\documentclass[a4paper,fleqn,usenatbib]{mnras}

\usepackage{newtxtext,newtxmath}

\usepackage[T1]{fontenc}
\usepackage{ae,aecompl}

\usepackage[dvipdfmx]{graphicx}	
\usepackage{amsmath}	
\usepackage{amssymb}	
\usepackage{multicol}
\usepackage{siunitx}
\usepackage{bmpsize}
\usepackage[anythingbreaks]{breakurl}
\usepackage{subfig}
\usepackage{fancyvrb}
\usepackage{hyperref}

\def\startdata{\if@table@not@headed\kill\caption{\\%
    \@tablecaption}\endhead\hline\endfoot%
  \fi%
}

\def\enddata{%
 \crcr 
 \noalign{\vskip .7ex}%
 \before@enddata 
 \endtabular 
 \setbox\pt@box\lastbox 
 \pt@width\wd\pt@box\box\pt@box 
}%

\newcommand{\aprx}{\raise.17ex\hbox{$\scriptstyle\sim$}}

\title[SPIDERMAN code package]{SPIDERMAN: an open-source code to model phase curves and secondary eclipses}

\author[T. Louden, L. Kreidberg]{Tom Louden$^{1}$\thanks{E-mail: t.m.louden@warwick.ac.uk} and Laura Kreidberg$^{2,3}$\\
$^{1}$Department of Physics, University of Warwick, Coventry, CV4 7AL, UK\\
$^{2}$Harvard-Smithsonian Center for Astrophysics, Cambridge, MA 02138, USA
$^{3}$Harvard Society of Fellows, Cambridge, MA 02138, USA}

\date{Accepted XXX. Received YYY; in original form ZZZ}

\pubyear{2016}

\begin{document}
\label{firstpage}
\pagerange{\pageref{firstpage}--\pageref{lastpage}}
\maketitle

\begin{abstract}

We present \textsc{spiderman}, a fast code for calculating exoplanet phase curves and secondary eclipses with arbitrary surface brightness distributions in two dimensions.

Using a geometrical algorithm, the code solves exactly the area of sections of the disc of the planet that are occulted by the star. The code is written in C with a user-friendly Python interface, and is optimised to run quickly, with no loss in numerical precision. Approximately 1000 models can be generated per second in typical use, making Markov Chain Monte Carlo analyses practicable. The modular nature of the code allows easy comparison of the effect of multiple different brightness distributions for the dataset.

As a test case we apply the code to archival data on the phase curve of WASP-43b using a physically motivated analytical model for the two dimensional brightness map. The model provides a good fit to the data; however, it overpredicts the temperature of the nightside. We speculate that this could be due to the presence of clouds on the nightside of the planet, or additional reflected light from the dayside. When testing a simple cloud model we find that the best fitting model has a geometric albedo of $0.32  \pm0.02$ and does not require a hot nightside. We also test for variation of the map parameters as a function of wavelength and find no statistically significant correlations.

\textsc{spiderman} is available for download at \url{https://github.com/tomlouden/spiderman}.

\end{abstract}




\section{Introduction}\label{sec:introduction}

Secondary eclipses and phase curves can give direct insight into the climate of exoplanet atmospheres. The details of heat transport in the atmosphere will affect how efficiently energy can be transported between the two hemispheres of the planet, controlling the day-night temperature contrast, and persistent wind patterns can move the brightest point of the planet from the sub-stellar point. As different wavelengths probe different depths of the atmosphere, spectroscopic phase curves and eclipses also contain information on the vertical temperature structure.

General Circulation Models (GCMs), are currently the state of the art for predicting the wind patterns and temperature distributions of the surface of exoplanets \citep[e.g.][]{Showman2008}. An important early prediction of these GCMs was that hot, tidally locked Jupiters would display peak brightness significantly offset from the substellar point. The hotspot offset is the result of fast superotating equatorial winds in the atmosphere that arise from a coupling of planetary scale circulation and day night advection \citep{Showman2011}.

The high equilibrium temperature of hot Jupiters affords them contrasts with their parent stars of order 0.1\%, making direct detection with current generation instruments feasible. \citet{Deming2005} detected the infrared flux of HD\,209458b by observing the secondary transit of the planet with the MIPS instrument on \emph{Spitzer}. Using \emph{Spitzer}/IRAC, \citet{Knutson2007b} went on to measure the secondary eclipse of HD\,189733b with high precision, and also show that the system displayed a sinusoidal \emph{phase curve}, caused by the bright regions of the planet rotating into and out of view. Intriguingly, the phase of this sinusoidal curve was significantly offset from the time of secondary eclipse, validating earlier predictions from GCMs. This effect can also be mimicked by a high eccentricity, but \citet{J.deWit2012a} show that this is not the case for HD\,189733b.

There appear to be two broad classes of exoplanet atmosphere heat circulation. Hot Jupiters like HD\,189733b, with relatively low temperatures have highly significant hotspot offsets and efficient day-night redistribution \citep{Knutson2007b}. In the other class, $\mu$ Andromedae, HD\,179949b and HD\,209458b are all significantly hotter than HD\,189733b and display much higher temperature contrasts between the day and nightsides, implying inefficient redistribution of energy. \citep{Harrington2006,Cowan2007,zellem2014}. This behavior can be explained with the results of General Circulation Models \citep{Komacek2015,Komacek2016}

The high velocity winds transporting this energy have been predicted to be on the order of kilometers per second, which would leave a measurable imprint on the spectral absorption lines in the planet's atmosphere through Doppler shifting \citep[e.g.][]{Showman2013}, and indeed, this effect has been observed in the transit of HD\,209458b \citep{Snellen2010} and HD\,189733b \citep{Louden2015}. 

As well as longitudinal mapping with phase curves, it is possible to get a two-dimensional map of the dayside of an exoplanet from the secondary eclipse, provided that there is a non-zero impact parameter \citep{Rauscher2006}. This technique works because the inclination of the system results in the occulting limb of the star scanning across the disc of the planet in both longitude and latitude during ingress and egress. The brightness distribution of the planet's permanent dayside will be imprinted on the precise shape of the ingress and egress curves. This has been demonstrated for HD\,189733 b by \citep{Majeau2012} and \citep{J.deWit2012}.

When phase curves are spectrally resolved, they can provide an extremely powerful tool for accessing the vertical structure of an exoplanet atmosphere as a function of longitude. \citet{Stevenson2014} use the Wide Field Camera 3 instrument on the Hubble Space Telescope \textsc{HST}/WFC3 to measure the emission spectrum and recover the temperature-pressure profile of WASP-43b as a function of phase.

An open-source code for calculating both exoplanet eclipses and phase curves in a self-consistent way for arbitrary brightness distributions does not currently exist. \textsc{batman} \citep{Kreidberg2015a} is an open source code to model exoplanet primary transits with a fast C-based analytical integrator, making it suitable for use with Markov chain Monte Carlo (MCMC) analysis, and has found broad use in the exoplanet community. \textsc{spiderman} is designed to fulfill the same role for secondary eclipses and phase curves.

In the past, authors studying phase curves have generally used non-physically motivated models to fit the data, such as sine curves or orange slices \citep{Cowan2008}. Whilst these approaches are still useful, \textsc{spiderman} enables a \emph{direct} fit to the underlying brightness distribution, and an easy way to compare different models.

In this paper we first describe the model, and provide a brief overview of the code, and then an example of use on the hot Jupiter WASP-43b.

\section{The model}\label{sec:the model}

\subsection{Integration scheme}\label{sec:integrator}

In order to accurately calculate both the phase curve and secondary eclipse the code must integrate over the portion of the planet that is facing the observer and is not occulted by the star at each time step. In order to be general use and ``model agnostic'' the integrator cannot assume any specific properties or symmetries in the brightness distribution. This can be done numerically, by sampling a square grid of points on the planet's surface and summing those that are visible, but to achieve high precisions a large number of points would need to be sampled, which would be slow and computationally inefficient.

Instead, \textsc{spiderman} uses a novel integration scheme where the planet is divided into a relatively small number of regions, and the area of each region that is not occulted by the star is calculated exactly through geometry. The average surface brightness of each region is determined by the chosen model and multiplied by the visible area to get the total flux.

\begin{figure}
	\begin{center}
		\includegraphics[width=1.0\columnwidth]{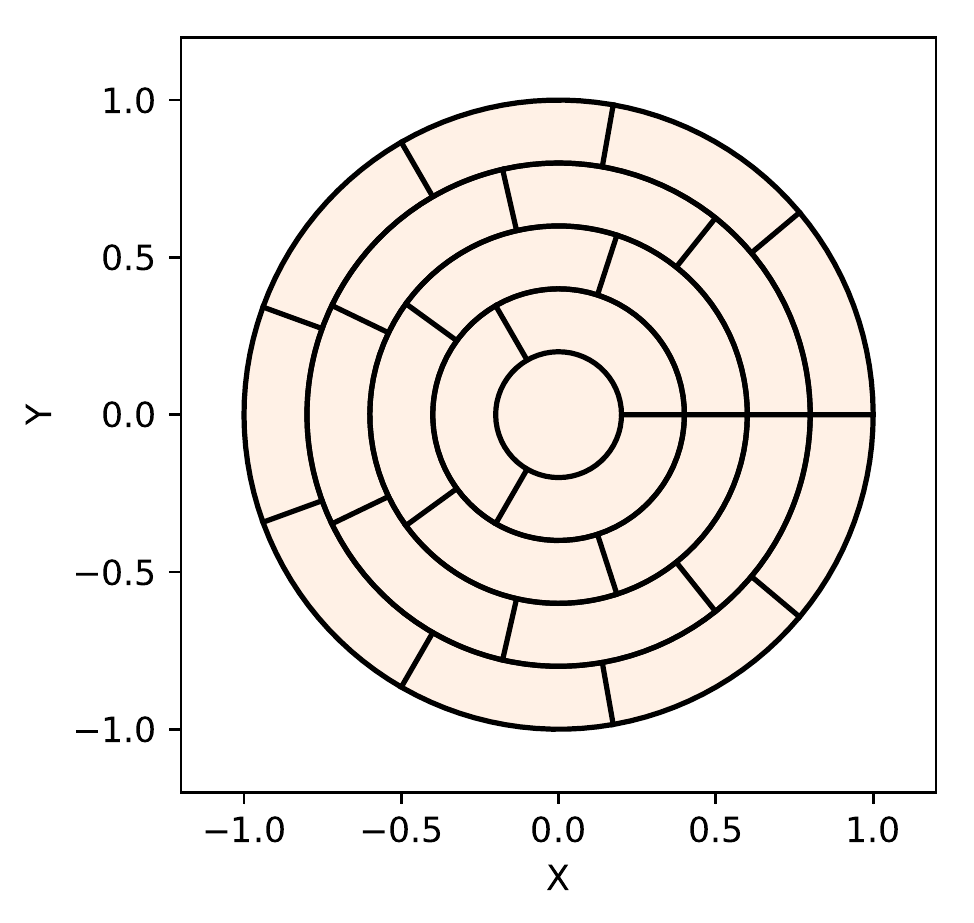}
		\caption{Schematic representation of the planet grid. All sectors have equal area. The co-ordinate scheme is defined as 0,0 at the centre of the planetary disc, and the planet is internally defined as having a dimensionless radius of 1.}
		\label{fig:schematic}
	\end{center}
\end{figure}

\textsc{spiderman} uses a radial co-ordinate system centered on the visible hemisphere of the planet. The planet is defined to have a radius of unity and the visible disc of the planet is divided into a series of annuli, and then these in turn are separated in polar angle $\theta$, as shown in the schematic in Figure \ref{fig:schematic}. The number of segments in each annulus are chosen so that the area of every sector is the same. This results in one circular segment in the center, then three regions in the second annuli and five in the third - with the number of segment in the $n$th annulus being $2n - 1$. Using this scheme, dividing the planet $n$ times radially results in a total of $n^2$ integration elements.

Each of the grid segments is a well defined geometric shape for which it is simple to calculate the fraction blocked by the planet passing behind the star, which in the model is represented as an occulting circle.

The area of one of these geometrical segments that has been blocked by the star can always be calculated by breaking the area into a combination of circle segments and triangles, for which the areas are known.

There are a small number of different general ``cases" for the numbers of triangles and segments that must be used, that can be selected based on the number and types of ``collision points" between the planet region and the stellar circle. Therefore, the first step for calculating the occulted area for each segment of the planet is to calculate these collision points. Within \textsc{spiderman} the region boundaries are stored as the radius of the inner and outer arcs, and the $\theta$ angle of the two radial lines.

Finding the collision points between an arc or a line and a circle (the star), if any exist, are trivial and fast to calculate. The code then classifies the geometric ``case" of the collision based on the number of collision points on the four boundary sides.

\begin{figure}
	\begin{center}
		\includegraphics[width=\columnwidth]{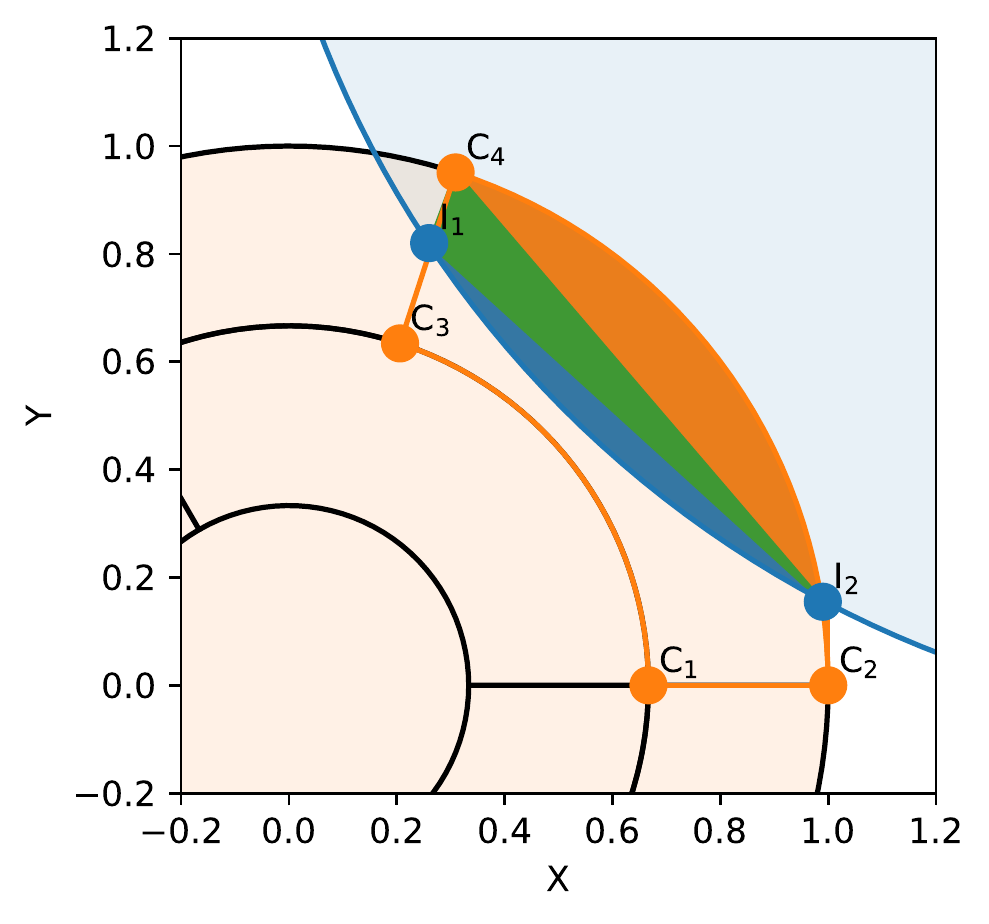}
		\caption{An example of the area calculation algorithm. Here the planet (light orange) passes behind the star (light blue). The occulted area of the highlighted region defined by the corners C$_1$,  C$_2$, C$_3$ and C$_4$ is calculated. The disc of the star intersects with the boundary of the region at I$_1$ and I$_2$. The overlapping area can be recovered exactly as the sum of two circle segments (blue and orange) and a triangle (green).}
		\label{fig:collision}
	\end{center}
\end{figure}

For example, if there are two collision points between the outer arc and the edge of the star, then this falls into the case where the occulted area can be described by the sum of two circular segments. Other cases include larger numbers of components, but can still be described by a small number of circle segments and triangles. A list of the major classes are given below.

An individual region is defined as having an inner and outer circle edge, and a first and second straight edge, which is the radial line from the center of the planet. Definitions are given in terms of the intersection points with the stellar disc (I$_1$ and I$_2$ in \ref{fig:collision}), and the corners of the region (C$_{1-4}$ in \ref{fig:collision}). The two corners closest to the center of the planet are referred to as ``inner" and the other two as ``outer" corners.

An example of one geometric case is given below, the remaining 11 cases are listed in the Appendix.

\subsection{Case 1: Outer circle crossed once, one straight edge crossed once.}

This case takes two forms: either one corner is blocked by the star, or three are.

The case for when one corner is blocked is illustrated in Figure \ref{fig:collision}. The total blocked area is recovered from the sum of three elements:

\begin{equation} \label{eq:inner_outer}
A_{tot} = A_1 + A_2 + A_3
\end{equation}

$A_1$ is the circle segment of the outer circular edge of the planet between the intersection point and the outer corner blocked by the stellar disc. A$_2$ is the the circle segment of the stellar disc between the two intersection points. A$_3$ is the triangle formed by the two intersection points and the corner blocked by the disc. In Figure \ref{fig:collision}, areas A$_1$, A$_2$ and A$_3$ are coloured in orange, blue, and green, respectively.

If instead the case is flipped such that 3 corners are blocked:

\begin{equation} \label{eq:inner_outer}
A_{tot} = A_R - (A_1 + A_2 - A_3)
\end{equation}

$A_R$ is the total area of the region. $A_1$ is the segment of the outer circle edge defined by the corner that is outside of the stellar disc and the intersection point. $A_2$ is the triangle defined by the two intersection points and the corner that is not blocked by the disc. $A_3$ is the circle segment defined by the stellar disc and the two intersection points.

While the calculation of overlapping errors is correct to within machine precision, numerical errors can be introduced by this approach due to averaging if the brightness distribution changes significantly faster than the area elements. The size of the elements can be adjusted as necessary by the user, at the cost of computation time.
In practice, we found that dividing the planet into five radial segments (25 total elements) gave sufficient precision for the smoothly varying brightness profile in our WASP-43b test case, and increasing the mesh density beyond this point did not change the results. Other grid schemes could also be defined by the user. For example, if the information content is higher in some parts of the planet than others, then the grid could be made finer in these regions.

We performed a simple validation check to demonstrate that the integrator is working correctly. Assuming a uniform brightness distribution, the number of sections that the planet is broken into for the integration should have no impact on the final light curve.

To test that the integrator is performing correctly we compare the case of a single segment model, i.e., a circle, for which there is a single analytical solution to the area blocked by another circle, to one with 400 segments. In both cases the planet is assumed to have a total luminosity that is 1\% that of the star.

\begin{figure}
\begin{center}
\includegraphics[width=\columnwidth]{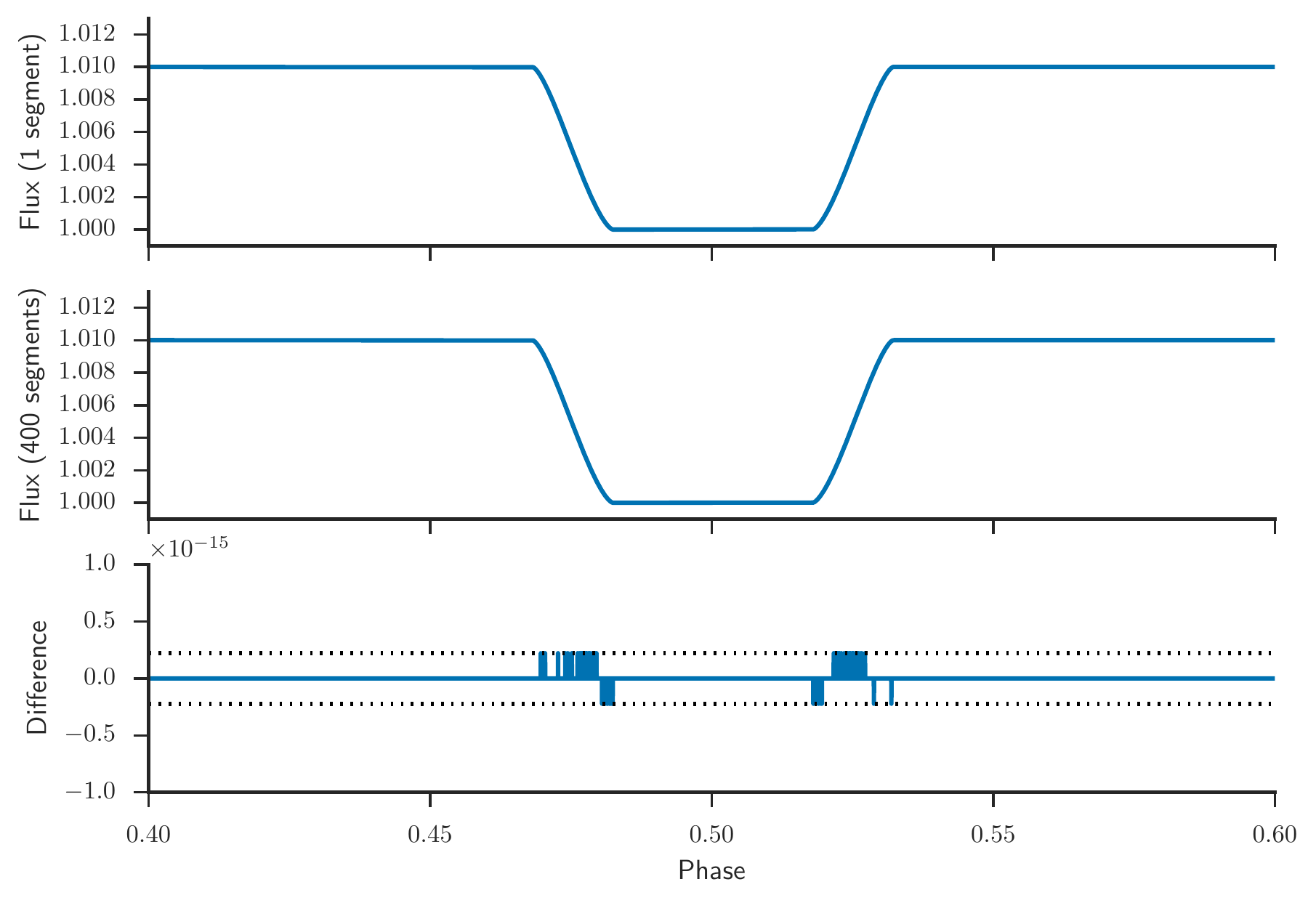}
\caption{A comparison between a uniform flux distribution eclipse model generated with 400 segments and a single segment. If the integrator is correct, then there should be no difference between these two models. The bottom panel shows that the results are identical, apart from errors at the level of floating point precision (dotted lines) during ingress and egress.}
\label{fig:precision}
\end{center}
\end{figure}

A light curve is calculated for each of these cases, and then the difference is taken. The results of this test can be seen in Figure \ref{fig:precision}, where it is clear that there is no difference between the two light curves greater than floating point precision (2.22e-16). We therefore consider the geometric integrator of \textsc{spiderman} to be validated.

\subsection{Brightness models}\label{sec:temp model}

\textsc{spiderman} has been written in a modular way, so that users will be able to choose from a variety of models to fit to their data, or provide their own. The currently available models range from fully non-informative to basic physical models. We also provide utilities to allow the results of forward models to be quickly converted into phase curves and eclipses. \textsc{spiderman} projects the model temperature or brightness map onto the visible sphere of the planet, and then directly calculates the secondary eclipse and phase curve. The option also exists to include an additional \emph{planetary} limb darkening component to the brightness distribution, which is parameterised as a quadratic law. The major currently available models are listed below. This list is not exhaustive, as it is intended that new models will be added to \textsc{spiderman} over time.

\subsubsection{Spherical harmonics}

\begin{figure}
\begin{center}
\includegraphics[width=\columnwidth]{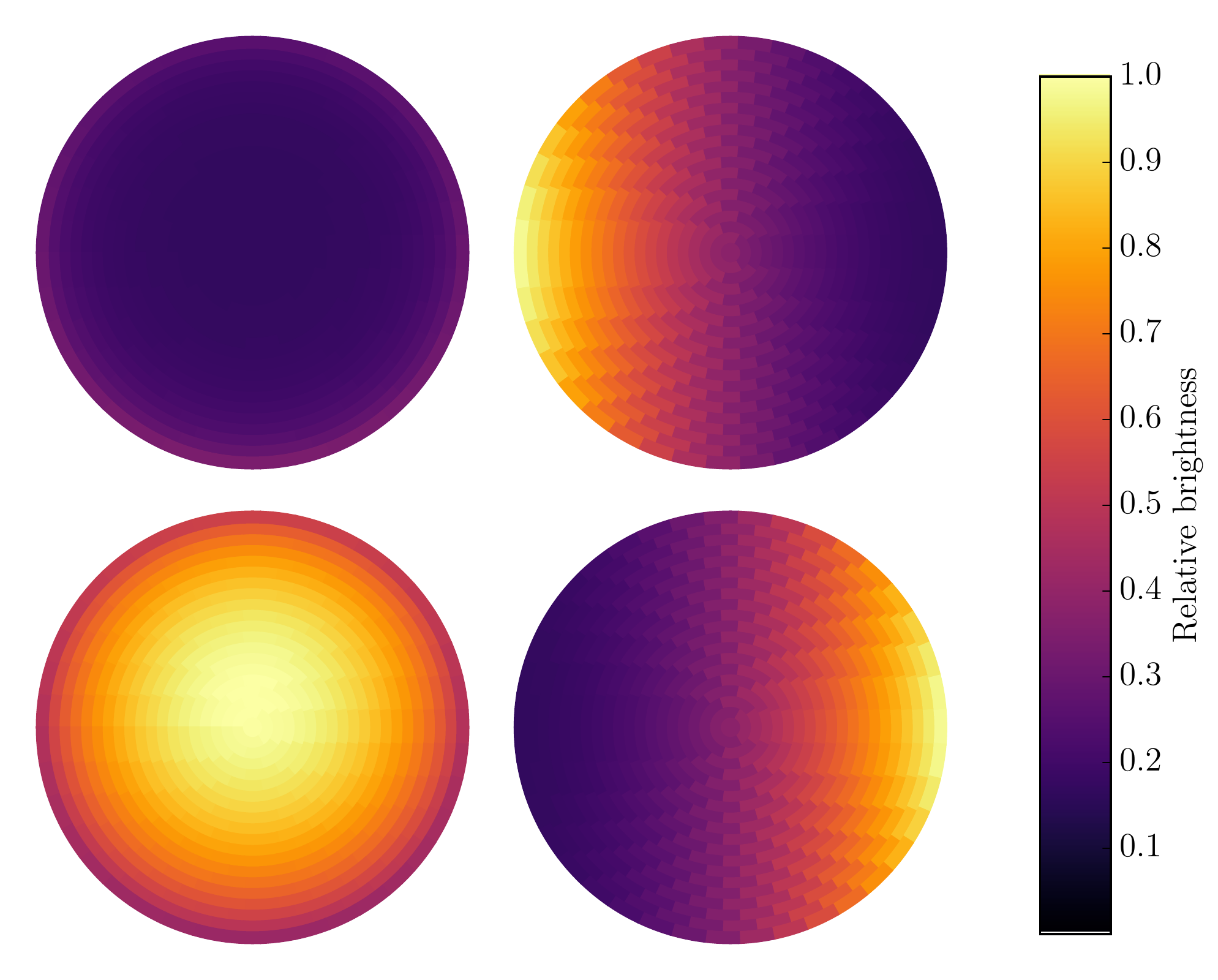}
\caption{An example \textsc{spiderman} model output generated using spherical harmonics. The planet is displayed as it would appear at 4 phases, clockwise from top left: 0.0,0.25,0.75,0.5.}
\label{fig:harmonics}
\end{center}
\end{figure}

A useful and physics-independent model is a sum of spherical harmonics. This method was used for the case of the phase curve of HD\,189733b by \citet{Majeau2012}. An example map generated by \textsc{spiderman} is displayed in Figure \ref{fig:harmonics}. The main observational features of a phase curve, including the offset hotspot, can typically be recovered with a only the first spherical harmonic, with the centre offset from the substellar point. \citep{Cowan2016} explore the effects of odd harmonics in phase curve data, and find that these can correspond to weather features in the planet atmosphere. \textsc{spiderman} uses a standard geodesy normalisation for calculating the harmonics, and can output maps in brightness or in temperature.

\subsubsection{Hotspot models}

The simplest physical model that one can construct involves defining a bulk planetary temperature or flux, and then a ``hotspot" where the size, contrast, and offset from the substellar point in longitude and latitude are model parameters. This model can be used to represent a ``two-temperature" planet with different fluxes on the dayside versus the nightside. The hotspot model could also be used to represent the reflection of light due to clouds in optical phase curves, which have been shown to be present in K2 lightcurves \citep{Demory2013} and are even found to be time varying in HAT-P-7 \citep{Armstrong2016}.

\subsubsection{Physical model}

We implement the kinematic model from \citet{Zhang2016}, which is a simple analytic function that closely replicates the temperature distributions of GCMs with only three free parameters. These parameters are physically motivated: the temperature of the nightside of the planet, $T_n$, the difference in temperature between the day and nightside, $\Delta T$, and the ratio between the advective and radiative timescales in the atmosphere, $\xi$.

\begin{figure}
\begin{center}
\includegraphics[width=\columnwidth]{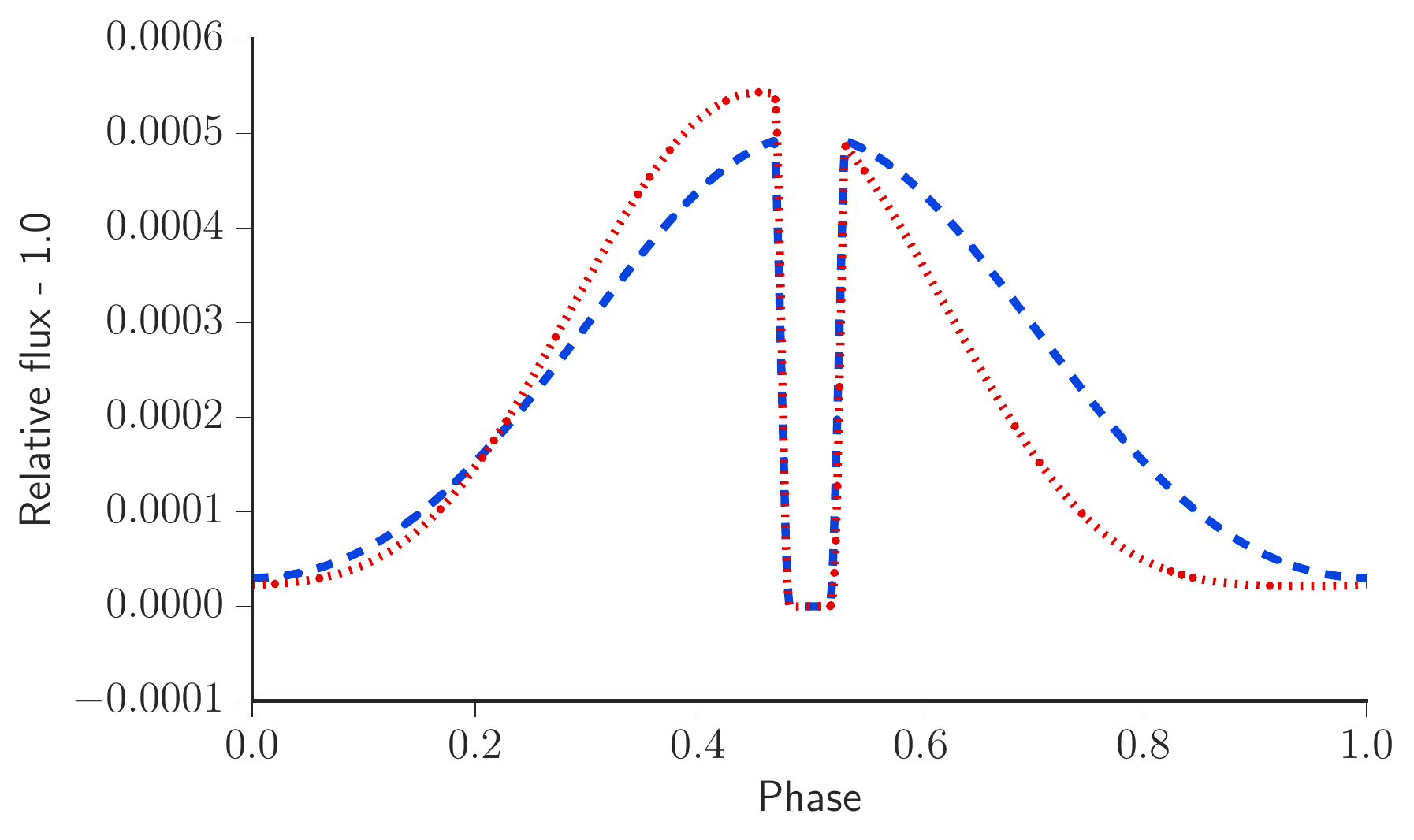}
\includegraphics[width=\columnwidth]{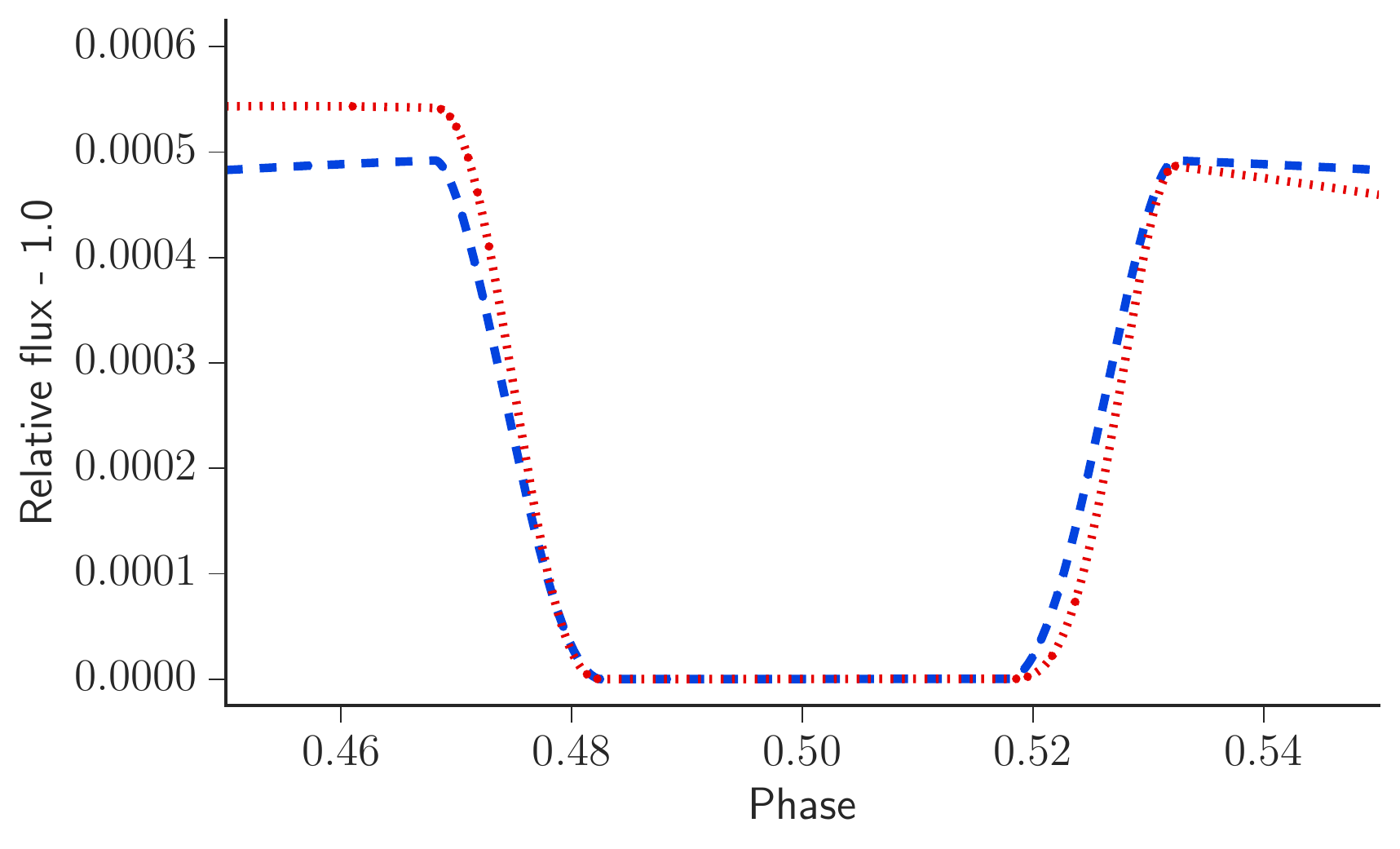}
\caption{The phase curve and secondary eclipses of an example planet generated with a simple spherical harmonics flux model (blue dashed) and the \citet{Zhang2016} physical model (red dot). Note in the physical model how the brightest point of the phase curve is significantly offset from the time of secondary transit, and the slightly different shape of the ingress and egress, both caused by the asymmetry in the flux distribution on the planet.}
\label{fig:ex_lcs}
\end{center}
\end{figure}

The $\xi$ parameter effectively parameterises the degree to which the hotspot of the planet will be offset from the substellar point by the circulation of the plant's atmosphere. The effects of this can be seen in \ref{fig:ex_lcs}, where both the position of the maxima of the phase curve, and the shape of the ingress and egress of the secondary eclipse are clearly affected by this asymmetry.

\subsubsection{Reflected light}

Most physical models assume that that the flux from a phase curve is nearly or entirely thermally emitted from the planet; however, in the optical and near-infrared a portion of the light will be reflected starlight. \textsc{spiderman} can use a stand alone reflection model, or combine it with a thermal emission model to get the total observable flux. The simplest model of reflection assumes a Lambertian sphere with a reflecting layer that reflects evenly at all phase angles, and can be characterized by a geometric albedo.

\subsubsection{Forward model}

\textsc{spiderman} also has the capability to quickly calculate a phase curve and secondary eclipse from a pre-calculated forward model. The code can read in a grid of either temperatures or brightnesses, in longitude and latitude, and will use a two dimensional bicubic interpolation within this grid to assign brightnesses to the grid points at each time step. Continuity is assumed east-west and across the poles to minimise edge effects.
This is significantly slower than the other models due to the interpolation scheme, but it is assumed that the generation of the forward model is by far the biggest bottle-neck in the calculation, so in this instance speed is not the priority.
This mode is intended to be used in particular for modelling eclipse mapping observations from 3D GCM calculations, either for comparison to data or to test analytical theories.

\subsubsection{Converting temperature to fluxes}\label{sec:ttof}

Some physical models, such as the one adapted from \citep{Zhang2016}, are defined in terms of temperature, so \textsc{spiderman} must convert that to a flux. To do so we make the assumption that the temperature corresponds to the \emph{brightness temperature} of the planet's photosphere. This will be valid for band passes that are narrower than strong molecular bands in the spectrum, such as water or carbon monoxide, but will not hold over wide band-passes.

There is not an analytical expression for the total amount of flux emitted by a blackbody between two arbitrary wavelengths, so \textsc{spiderman} calculates this value numerically. A grid of blackbody curves multiplied by the instrument response is pre-calculated as a function of temperature, and the flux in the chosen wavelength region is summed numerically. \textsc{spiderman} then linearly interpolates on this grid when assigning fluxes to the planet integration regions based on their temperature.

\subsection{Stellar model}\label{sec:stellar model}

\textsc{spiderman} models the stellar spectrum using a grid of PHOENIX stellar atmospheres calculated by \citet{Husser2013}. It is also possible to use alternative models, or a simple blackbody spectrum, which will typically be sufficient for broadband observations. Using the user provided wavelength bandpass and instrument response, \textsc{spiderman} calculates the total observed flux for the stellar models as a function of effective temperature, it then interpolates on this grid using a spline.

Since the shape and depth of an eclipse and phase curve are determined by the \emph{ratio} of the fluxes between the two bodies, there will clearly be an exact degeneracy between the planet and star brightness. In order to recover physical parameters of the planet, such as temperature, it is necessary to place constraints on the \emph{absolute} scale or brightness of the system. These constraints only need to be on flux as a function of surface area, i.e., on the specific intensity of the system components, and on their radius ratio, not on their total luminosity or radii.

$(R_p / R_*)^2$ is well constrained from primary transits - the small variation as a function of wavelength is of little importance. There are usually good measurements of the effective temperature of the star from spectroscopy and stellar modelling. These additional parameters fix the flux scale of the system, allowing useful constraints to be placed on the absolute surface flux and brightness temperature of the planet.

\subsection{Light curve calculation}\label{sec:lightcurve}

The primary purpose of \textsc{spiderman} is to produce light curves to compare with the data. Now that the functioning of the geometric integrator has been described, light curve generation is quite simple.

Given a series of points in time, the ephemeris of the system, the system scale and the planet/star radius ratio, the position of the planet relative to the star can be calculated for each timestep. Light time of travel effects are accounted for, with the semi-major axis in physical units as a parameter.

Since we assume that the planet is tidally locked, at each time step the geometric grid representing the planet is populated with the appropriate fluxes from the chosen flux distribution model.

Then, for each segment on the planet, it is checked whether it is currently being occulted. This is quickly achieved by checking for contact between the edges of the region and the edge of the star. If there are no contact points, and the distance from the center point to the center point of the star is greater than the stellar radius, then the segment is fully visible, and the full flux is counted. If the distance is less then the stellar radius, then it is fully occulted and no flux is added. For cases of partial coverage, the previously described geometric integrator is used to calculate the flux that should be added.

To this, the pre-calculated stellar total flux is added to give the total system flux. This is then divided by the stellar flux to give a relative light curve that can easily be compared to data.

Note that \textsc{spiderman} does not calculate primary transits. To fit models to data that contain both primary and secondary eclipses, \textsc{spiderman} should be combined with another package such as \textsc{batman}. The parameter naming convention and call sequence from \textsc{spiderman} has been chosen to be compatible with \textsc{batman} for this reason. 

\subsection{Code availability}\label{sec:package}

\textsc{spiderman} is an open-source project and is being actively developed on github. The code is available on PyPI and github at \url{https://github.com/tomlouden/spiderman}. 
Installation instructions and documentation are available at \url{http://spiderman.readthedocs.io}.

\begin{figure*}
\begin{center}
\includegraphics[width=\textwidth]{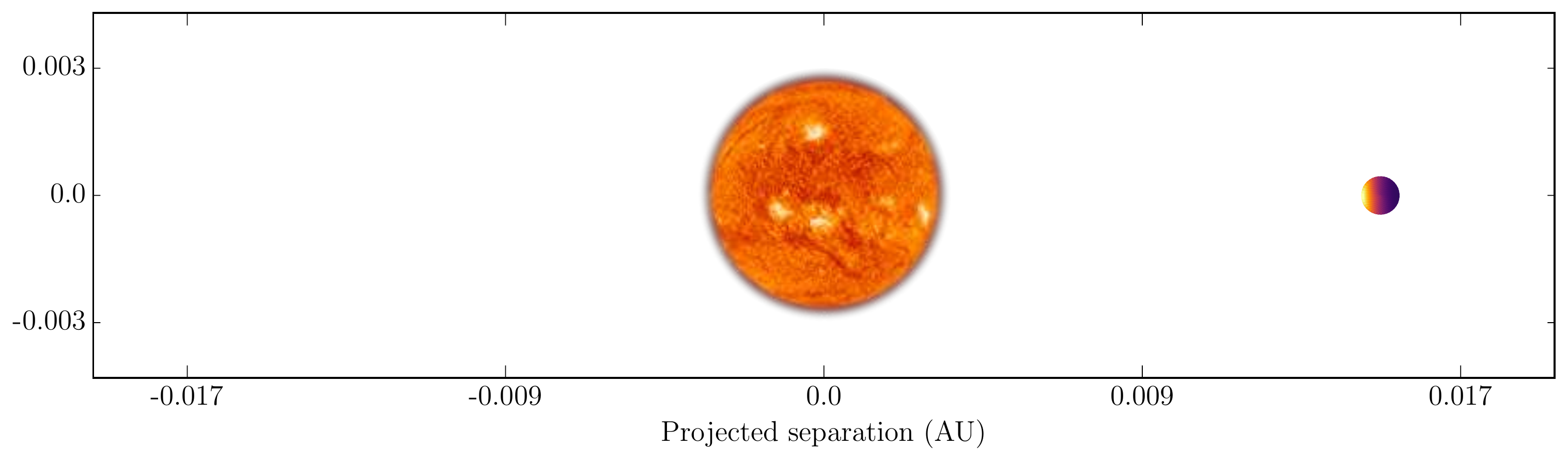}
\caption{An example of the graphical output that \textsc{spiderman} can produce. This image is a scale representation of the system, with the planetary flux map included. This is useful for visualising the model. It was generated with the spiderman.plot\_system() command. The image of the star is purely aesthetic and is not related to any model parameters (image credit Solar Dynamics Observatory).}
\label{fig:plot_system}
\end{center}
\end{figure*}

\begin{figure}
\begin{center}
\includegraphics[width=\columnwidth]{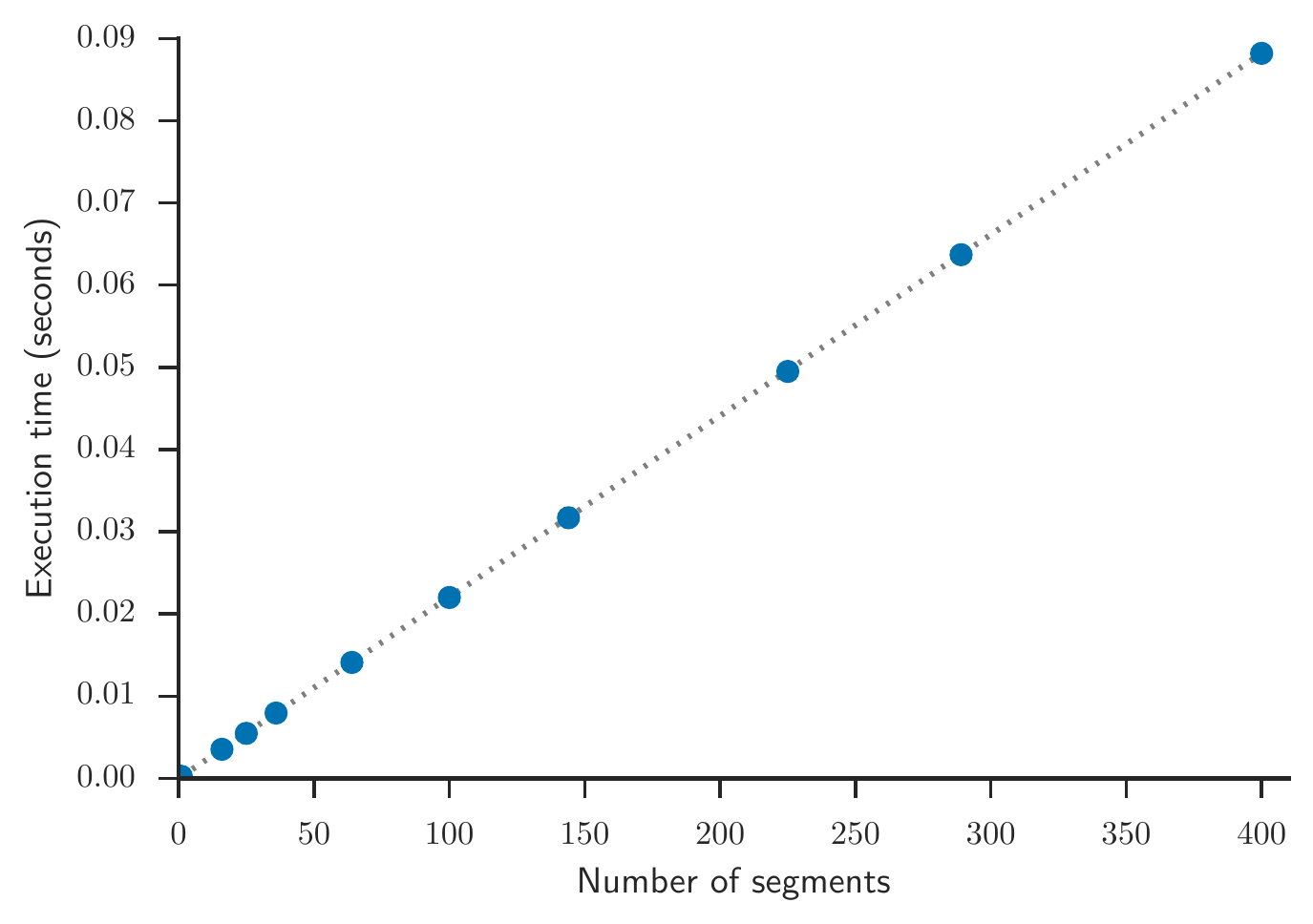}
\caption{The speed with which a model with 1000 timepoints can be generated increases linearly with the total number of segments in the model.}
\label{fig:exec_time}
\end{center}
\end{figure}

The code is written in Python with C extensions, and was optimized to run quickly. In typical usage, SPIDERMAN is capable of producing over 1000 models per second on a single core. This means that a 1 million MCMC sample can be generated in approximately a quarter of an hour. Figure \ref{fig:exec_time} shows the result of a performance test carried out on a single core of an Intel Core I5-3470 Processor, where we calculated the length of time to produce a model with 1000 time points as a function of the number of segments of the model. The test shows that the execution time scales linearly with the number of segments in the model, which is expected, showing the overhead to run the model is very small. The precision required for typical usage can easily be reached using 16 or 25 segments, but larger numbers of segments can be used to produce better looking diagrams. The diagrams in this paper were generated with a 400-segment model.

In addition to light curve generation, \textsc{spiderman} also includes plotting functions to visualise the brightness or temperature distributions. An example plot is given in \ref{fig:plot_system}, which is a scale diagram of the system at a specified time or phase. Plots like this are useful for visualising the effects of varying the parameters.

\section{Application to data}\label{sec:Application}

We tested \textsc{spiderman}'s performance for a phase curve and secondary eclipse of the hot Jupiter WASP-43b observed with \emph{HST}/WFC3. These data have previously been published in \citet{Stevenson2014}. However, in their paper the phase curve is fit with a sinusoid function. Here we instead apply the physically motivated model of \citet{Zhang2016} to test its applicability to observed data. This also reduces the total number of parameters needed for the fit. For our model fit we take the raw lightcurves from these observations and detrend them simultaneously with our model fit.

Planetary limb darkening is a potential sources of uncertainty on the final brightness distribution that would require more detailed atmosphere models to predict. \textsc{spiderman} has the option to include planetary limb darkening in addition to the chosen brightness model with a quadratic law. In principle, measurements of planetary limb darkening can constrain the vertical temperature profile of the planet; however, this would require a full radiative transfer treatment calculation of the atmosphere. For the case of WASP-43b, where the limb darkening is unknown, we initially fix the limb darkening to zero, and then test the impact of varying the limb darkening, treating the additional freedom as a test of the sensitivity to uncertainties of the radial brightness profile of the planet.

\subsection{Systematics model}\label{sec:systematics}

As with most exoplanet observations with \emph{Hubble}, the data are affected by instrument-based systematic errors. Fortunately, the sources of these errors are relatively well understood, and are generally repeatable between observations. They are therefore straightforward to fit for and remove. 

The systematics have effects on scales of \emph{visits}, \emph{orbits} and individual exposures, where a visit is a continuous set of orbits directed at a target. Due to the low Earth orbit of \emph{Hubble}, each complete orbit lasts 96 minutes, and for approximately half of this time the target will be unobservable because it is behind the Earth. For this reason, a complete exoplanet transit can not be observed with one ``visit", and it is necessary to observe several transits with different temporal phasing to completely cover the full orbital phase of the system. In this study, three visits, each with 13 or 14 orbits was used to cover the full phase of WASP-43. The phasing can be seen in Figure \ref{fig:systematics}, and a phase-folded lightcurve can be seen in Figure \ref{fig:phase folded}

The systematics model we use to fit WASP-43 is the same that was used in the previous treatment \citep{Stevenson2014}. There are three components to the systematics model, the scan direction offset, ``hook effect'' and visit long slopes. The scan direction offset depends on which way the the detector was scanning during the readout. In one half of exposures the scan direction is in the same direction as the readout, and in the other half it is the opposite. This leads to a very repeatable offset in the fluxes between the two scan directions, which can be captured with a single parameter \emph{scale}

Each orbit displays a characteristic exoponential ``ramp'' signal, which can be modelled as
\begin{equation} \label{eq:hook}
F_{cor} = F*(1-\exp(-r1 t_\mathrm{orb}-r2-r3 \delta_\mathrm{fo}))
\end{equation}
Where $F$ is the flux before correction and $F_{cor}$ is the corrected flux $t_{orb}$ is the time from the start of each \emph{orbit}, $r1$, $r2$, $r3$ are the fitted coefficients. This exponential ramp has a different shape during the first orbit of a visit, so the $\delta_\mathrm{fo}$ function takes a value of 1 if the orbit is the first, and 0 at all other times. The exponential ramp shape is believed to be related to charge trapping, and is a characteristic and repeatable feature across orbits, so these parameters are fit simultaneously over visits.

Each \emph{visit} displays a decreasing slope, which is modelled as a second order polynomial. The cause of these features is unknown, and they have not been linked to any physical parameters of the telescope \citep{Wakeford2016}. The shape of these visit-long effects is slightly different between the three visits used here, so the three parameters of the polynomial are fit seperately for the three visits.

The feature that is the greatest cause for concern is the visit-long slope in the data, as this has a comparable timescale and amplitude to the planet's phase curve, so there is potential for degeneracies between the systematics and physical models. Fortunately, the phasing of the observations is such that the three visits cover different portions of the phase curve, which helps to alleviate the degeneracy.

For \emph{HST} systematics, where the major sources of error are fairly well understood, a parameterised systematics model of this form works well \citet{Wakeford2016}. Alternatively, detailed instrumental models can recover the same features with fewer free parameters.  However, if the form of the systematics changes over time then this approach might underestimate the uncertainty in the shape of the systematics and thus artificially constrain the planet models. An alternative, powerful technique which does not make assumptions about the form of the systematics is Gaussian Processes \citep[e.g.][]{Gibson2012a}.

\subsection{MCMC}\label{sec:MCMC}

In order to determine the errors on the parameter values of interest and test for degeneracies with the systematics model, we perform an MCMC analysis. We use the \textsc{emcee} affine invariant Monte Carlo implementation of \citet{Foreman-Mackey2013}.

The initial parameter positions are found with a least squares minimization approach, and the 1000 walkers were initialized around these best fitting values. For the orbital and system parameters we use published values and error bars as priors. For the period, inclination and $a/R_*$ we use the results of \citet{Hoyer2016}, and for $a$ the value from \citet{Hellier2011a}, The eccentricity was fixed at zero, which is supported by the findings of \citet{Hoyer2016}. The stellar log g and effective temperature priors are from \citet{Gillon2012}, and we calculate the stellar fluxes using the PHOENIX model stellar atmospheres calculated by \citet{Husser2013}. We assume that the planet is tidally locked to the star in all model fits.

To test for the potential impact of \emph{planetary} limb darkening, we ran a separate MCMC with this effect modeled by a simple quadratic limb darking law with parameters $p_{u1}$ and $p_{u2}$. We place the constraint that the limb darkening must be monotonic across the disc of the planet, i.e., there can be no change in the sign of the gradient. This results in profiles that are either flat, wholly limb darkened or wholly limb brightened, with no overly flexible profiles.

For both models, the MCMC was run for 100,000 steps with 1000 walkers, with the first 10,000 steps discarded as burn-in. Convergence was checked by visual inspection of the chains for each model parameter.

\subsection{Results and discussion}\label{sec:results}

\begin{figure}
\begin{center}
\includegraphics[width=\columnwidth]{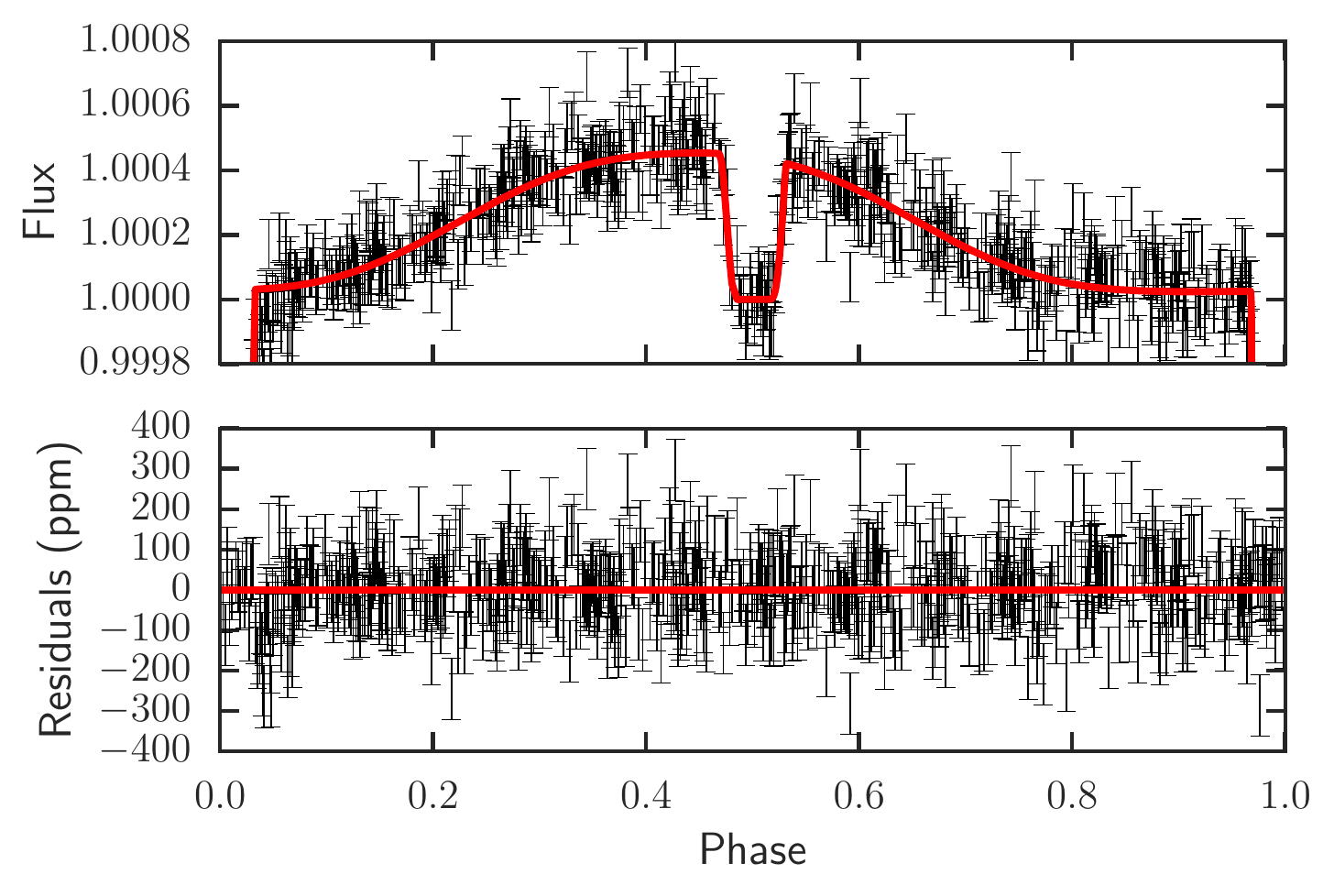}
\caption{The final fitted lightcurve using the \citep{Zhang2016} brightness model. The primary transit is fit by \textsc{batman} and is off the scale of this figure. The model flux immediately before and after primary transit is non-zero, indicating the presence of non-zero nightside flux.}
\label{fig:phase folded}
\end{center}
\end{figure}

\begin{figure*}
\begin{center}
\includegraphics[width=1.0\textwidth]{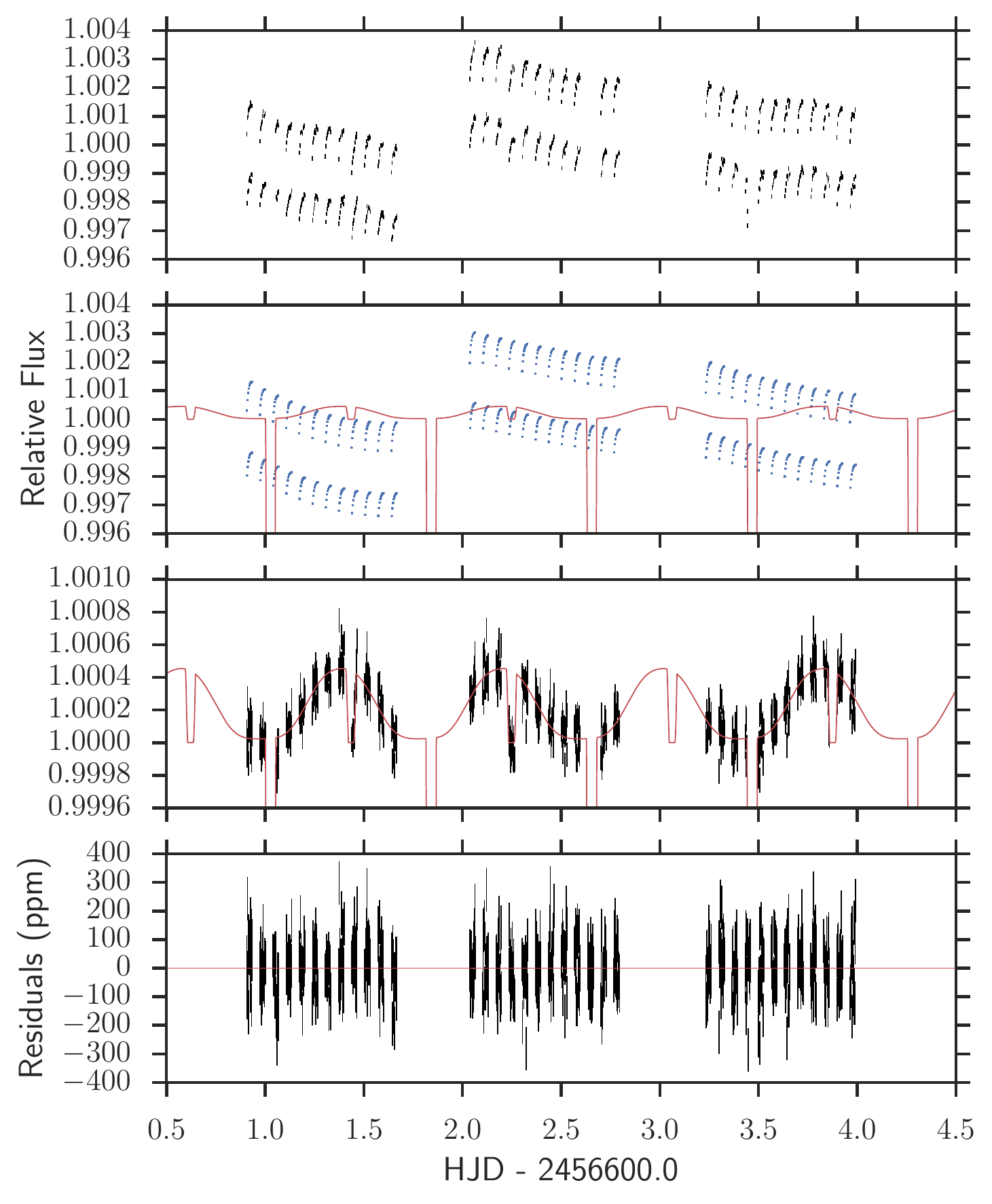}
\caption{Top: raw data from HST, the systematic components are all visible. second panel shows the physical model fit (red) and the systematic model (blue). The third panel shows the model fit to the data with the systematic noise component removed, the fourth panel is the residuals of the fit.
}
\label{fig:systematics}
\end{center}
\end{figure*}

The best fitting model and the residuals are shown in Figure \ref{fig:phase folded} as a function of planetary phase, after correction for instrument systematics. The data and model are also plotted as a function of time in Figure \ref{fig:systematics}. To demonstrate the systematics model the data are plotted with and without systematics correction.

\begin{figure}
\begin{center}
\includegraphics[width=\columnwidth]{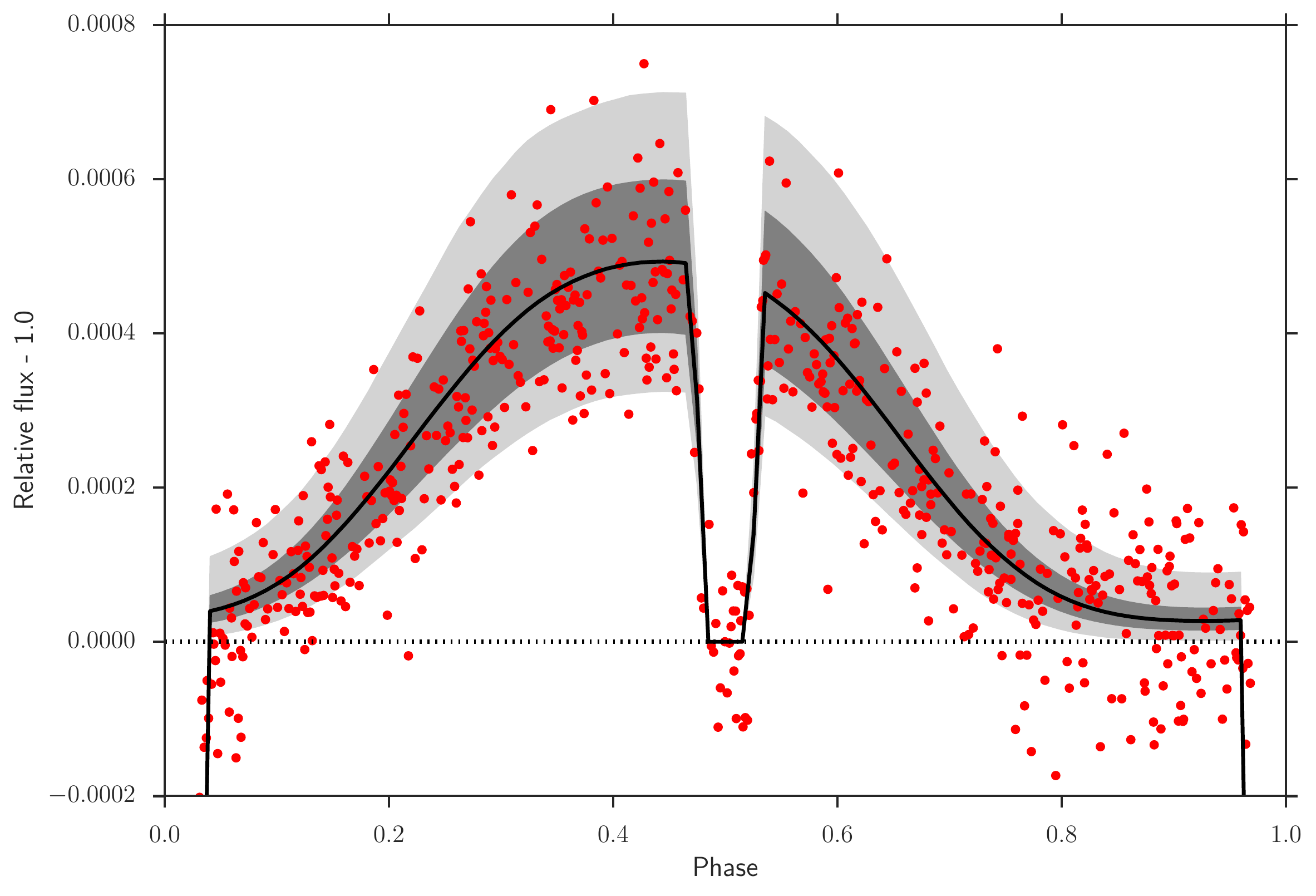}
\caption{The uncertainty on the model from the MCMC fit not of the Zhang model including clouds or planetary limb darkening. The dark grey region is the 1 $\sigma$ confidence region, light grey is 3 $\sigma$. The implied flux from the nightside is greater than 1 at the 3 $\sigma$ level. The apparent non-zero nightside flux is model dependent due to degeneracies with the systematics detrending and is not significant in the other two models tested.}
\label{fig:model}
\end{center}
\end{figure}

The model fit implies that the nightside of the planet is significantly hotter than expected at 1290 $\pm 39$ K, whereas it is not significantly detected in \citet{Stevenson2014}. To establish the significance of this result, we test whether the flux immediately before and after primary transit is significantly different to the star-only flux visible during the secondary eclipse. Drawing 10,000 lightcurves from the parameters in the non-limb darkened MCMC posterior, we plot the median, 1 $\sigma$ and 3 $\sigma$ contours for the lightcurve in Figure \ref{fig:model}. The flux is found to be greater than 1.0 at the 3 $\sigma$ level.

\begin{figure*}
\begin{center}
\includegraphics[width=\columnwidth]{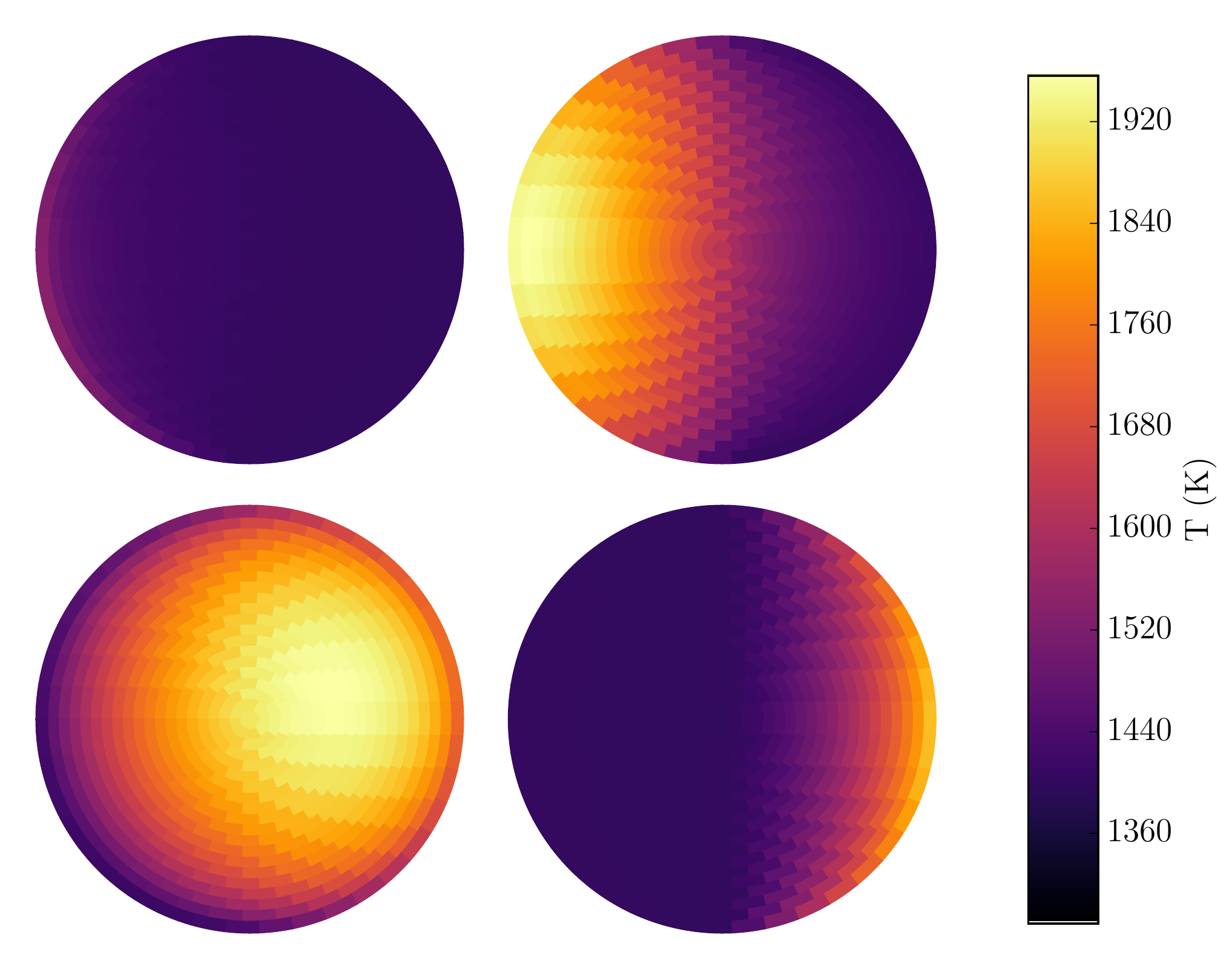}
\includegraphics[width=\columnwidth]{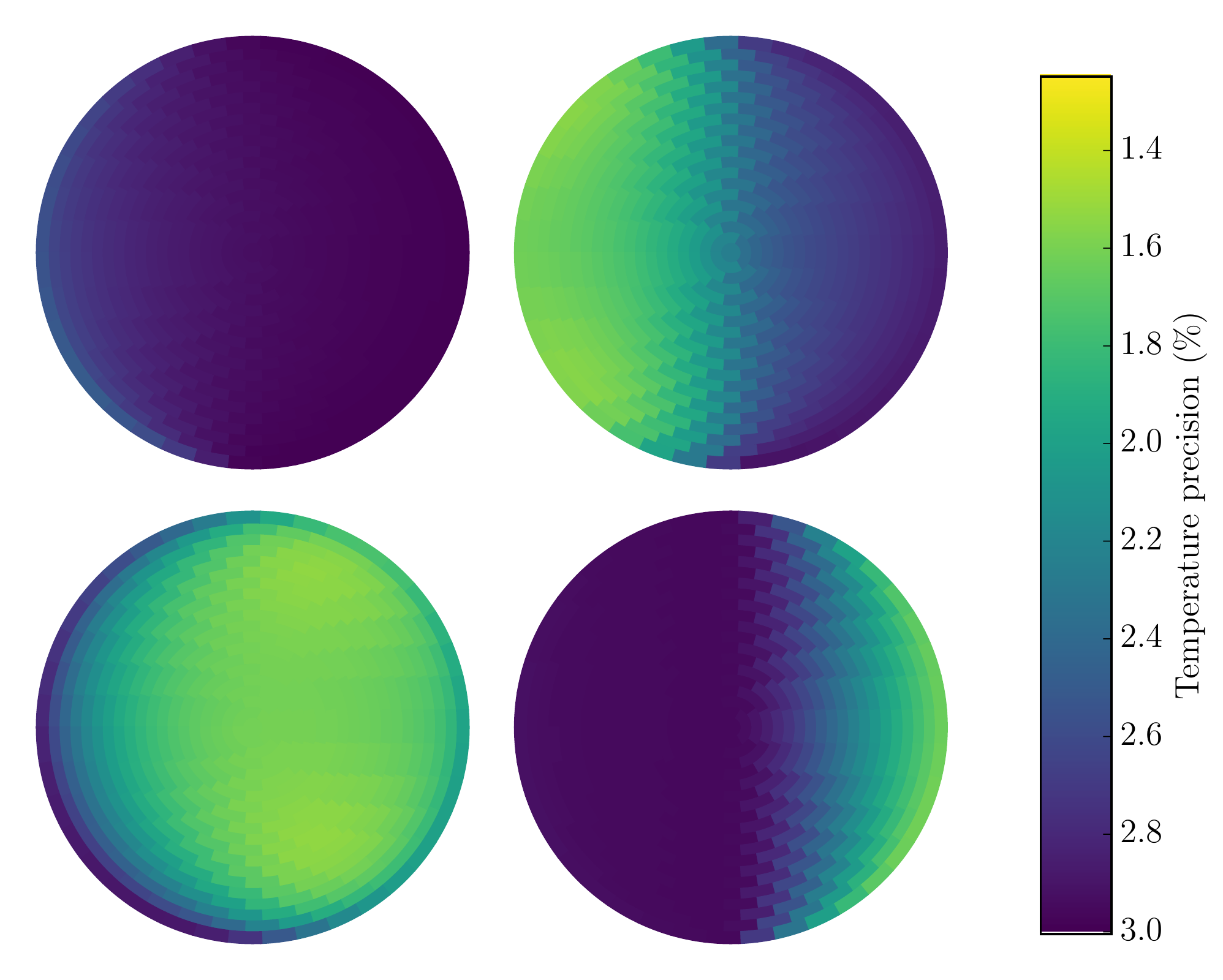}
\caption{The underlying temperature distributions generated by the Zhang model corresponding to figure \ref{fig:best_fit_flux} with the precision attained.}
\label{fig:best_fit_temp}
\end{center}
\end{figure*}

The visual representation of the flux maps for the non-limb darkened case can be seen in Figure \ref{fig:best_fit_flux}. The best fitting model is displayed at four phases, as well as the relative error on the flux distribution. The error map was generated by drawing 10,000 maps from the MCMC posterior and taking the standard deviation of the fluxes in each segment. The offset hotspot is significant and clearly visible. The underlying temperature maps, and the errors in these, are shown in Figure \ref{fig:best_fit_temp}.

\begin{figure}
\begin{center}
\includegraphics[width=\columnwidth]{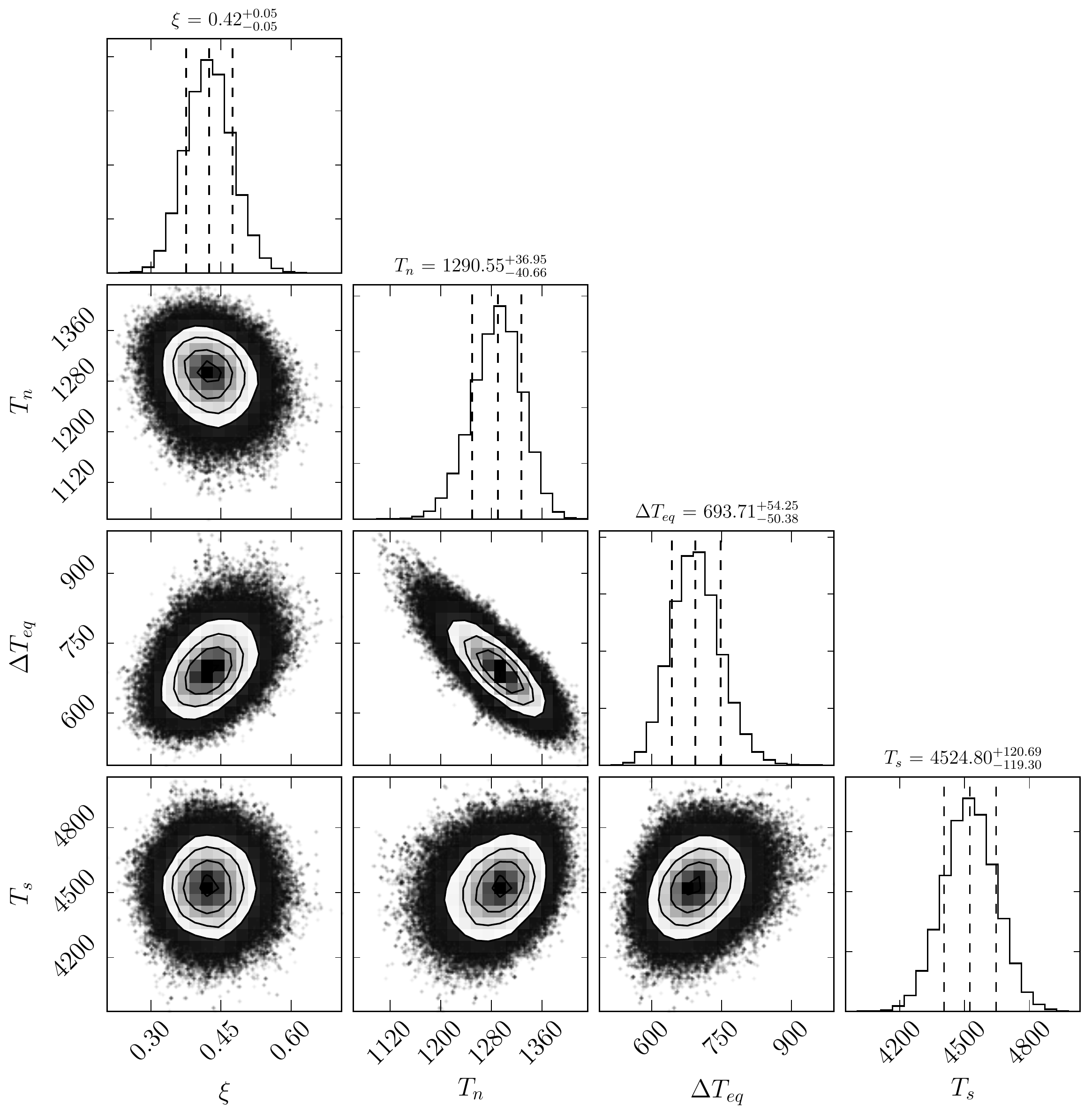}
\caption{A corner plot for the fixed limb darkening fit focusing on the key model parameters. The parameters for the Zhang model fit are all loosely correlated. The distribution for the stellar temperature, $T_s$, is set by the prior}
\label{fig:triangle corner sub}
\end{center}
\end{figure}

The correlation between the physical model with the other model parameters can be seen in figure \ref{fig:triangle corner sub}. $T_s$ and $\Delta T$ are clearly correlated parameters, since between them they determine the average temperature of the dayside and hence the amplitude of the phase curve. $\xi$ is not strongly degenerate with the other parameters. This indicates that the $\xi$ parameter is robust to any errors in the absolute flux level of the system.

The fit was repeated with limb darkening enabled to test the impact of the additional radial flexibility on the model. With the exception of $T_n$, which is lower by more than $3\sigma$ in the fit with planetary limb darkening, the physical parameters are consistent. In particular, $\xi$ is only weakly affected by the presence of limb darkening, the difference being much less than $1\sigma$. This is not unexpected, as $\xi$ effectively parametrises the asymmetry in the flux distribution, while the limb darkening is symmetrical. $\xi$ is directly tied to the details of heat transport in the atmosphere, so the insensitivity of this parameter to limb darkening is useful, as it shows that it can be recovered even with systematic uncertainties in the flux distribution, so long as they are symmetric.

That the fit with limb darkening enabled finds a significantly fainter nightside flux suggests that the nightside flux may be an artifact of using a simple physical model. This may be due to a correlation between model parameters making the fit inflexible and the simultaneous fit with the systematics model. Alternatively, the model may be unable to parametrize a feature of the nightside that is blocking the emitted flux, such as clouds.

Another possibility is that the apparently anomalously high nightside flux may be a consquence of the physical model missing some relevent atmospheric physics on the \emph{dayside}, which will artificially constrain the result. One potential factor is that the presence of clouds in the atmosphere could provide additional reflected light in addition to the thermally emitted light on the dayside of the planet. Since reflected light is not included in the model, and the parameters of the physical model are correlated, the best-fitting model may favour making the entire planet hotter in order to recover the dayside flux. \cite{Keating2017} investigated the impact of adding reflective clouds to the phase curve of WASP-43b, and found a surprisingly high near-infrared geometric albedo of $0.24 \pm 0.01$.

We add an additional very simple cloud model to demonstrate the potential impact of this effect on the model fit. We add an additional geometric albedo parameter for the planet, and for our toy model assume uniform cloud coverage and that the planet behaves as a Lambertian sphere. \textsc{spiderman} is capable of combining models, so the brightness from the reflected light is added to that emitted by the thermal model. The MCMC fit is repeated with this additional one free parameter and with planetary limb darkening set to zero.

We find that allowing a non-zero planetary albedo makes a significant difference to the model fit, and the emitted flux from the nightside becomes consistent with zero, with a 99.7\% upper limit for the nightside temperature of 860 K, significantly lower than the value of $1076 \pm 11$ K \cite{Keating2017} find using both WFC3 and \emph{Spitzer} data. The best fitting albedo is found to be $0.32 \pm 0.02$, which is somewhat higher than the best fit value of $0.24 \pm 0.01$ reported by \citep{Keating2017}; however, the results are alike in that they emphasize the potential importance of bright dayside clouds. The differences in our results are likely due to the differences in our thermal emission models and the difference in spectral coverages considered. \citet{Keating2017} only consider continuum emission of WASP-43b (excluding the water absorption feature), whereas we fit the broadband WFC3 phase curve.

We emphasize here that the ``temperatures" of the \citet{Zhang2016} model parameters do not correspond trivially to the average temperatures of the generated temperature map - for example, one cannot take the addition of the $T_n$ and $\Delta_T$ parameters as a representative dayside temperature for the model. When comparing the average temperature of the maps we will use the brightness temperature of a uniform sphere that produces the same total flux, to remove ambiguity. 

The addition of the reflecting clouds significantly changes the parameters of the temperature distribution, with the dayside also becoming correspondingly cooler, equivalent to an average dayside brightness temperature of 1220 K, compared to 1640 K in the model that did not include reflective clouds. 

The ratio of the radiative to advective timescales is increased to $2.2 \pm  0.5$, corresponding to a more significantly offset hotspot, though due to the combination of the emitted and reflected light the \emph{brightest} point of the planet in this bandpass is in the same place as the other models.

The results of all three runs are presented in Table \ref{tab:results}

\begin{center}
\begin{table*}{
\caption{The full results from the MCMC fit for the three models considered in detail: with planetary limb darkening fixed to zero, with planetary limb darkening as additional free parameters, and with reflective dayside clouds parameterized as a lambertian sphere with a geometric albedo. Note that while the parameters for the model of \citep{Zhang2016} have units of K they should not be interpreted as average values for the day and nightside of the planet. We report the mode and 1 $\sigma$ Highest Probability Density region for all parameters: this is identical to the median and 1 $\sigma$ percentiles except for the $T_n$ parameter in the reflecting cloud case, which has a highly asymmetric posterior that approaches 0.}
\begin{center}
\begin{tabular}{l c c c c c c c}
\hline
\hline
          &   ld fixed   &             &   ld free    &             &   clouds     &             \\
name      &   median     & error       &   median     & error        &   median     & error        & unit        \\
\hline
Period & 0.813473977 & $\pm$0.000000036 & 0.813473978 & $\pm$0.000000034 & 0.813474000 & $\pm$0.000000030 & days\\
$t0$ & 2456601.0274024 & $\pm$0.0000091 & 2456601.0274023 & $\pm$0.0000092 & 2456601.0274048 & $\pm$0.0000090 & days\\
$a/R_*$ & 4.8762 & $\pm$0.0089 & 4.8770 & $\pm$0.0086 & 4.8774 & $\pm$0.0089 & -\\
inclination & 82.123 & $\pm$0.038 & 82.124 & $\pm$0.037 & 82.125 & $\pm$0.037 & degrees\\
eccentricity & 0 & - & 0 & - & 0 & - & -\\
$R_p/R_*$ & 0.159692 & $\pm$0.000093 & 0.159650 & $\pm$0.000091 & 0.159620 & $\pm$0.000091 & -\\
$u1$ & 0.3880 & $\pm$0.0083 & 0.3875 & $\pm$0.0083 & 0.3875 & $\pm$0.0083 & -\\
$u2$ & 0 & - & 0 & - & 0 & - & -\\
$c_1$ & 3.687319e+08 & $\pm$0.000090e+08 & 3.687522e+08 & $\pm$0.000088e+08 & 3.68757e+08 & $\pm$0.000067e+08 & -\\
$c_2$ & 3.693661e+08 & $\pm$0.000097e+08 & 3.693653e+08 & $\pm$0.000098e+08 & 3.693776e+08 & $\pm$0.000093e+08 & -\\
$c_3$ & 3.689717e+08 & $\pm$0.000093e+08 & 3.68984e+08 & $\pm$0.00010e+08 & 3.689801e+08 & $\pm$0.000073e+08 & -\\
$v1_1$ & -1.484e+06 & $\pm$0.045e+06 & -1.558e+06 & $\pm$0.048e+06 & -1.533e+06 & $\pm$0.044e+06 & -\\
$v1_2$ & -6.31e+05 & $\pm$ 0.50e+05 & -6.18e+05 & $\pm$ 0.55e+05 & -6.74e+05 & $\pm$ 0.49e+05 & -\\
$v1_3$ & -8.56e+05 & $\pm$ 0.49e+05 & -8.56e+05 & $\pm$ 0.54e+05 & -8.01e+05 & $\pm$ 0.48e+05 & -\\
$v2_1$ & 1.004e+06 & $\pm$0.059e+06 & 1.076e+06 & $\pm$0.063e+06 & 1.027e+06 & $\pm$0.058e+06 & -\\
$v2_2$ & 1.83e+05 & $\pm$ 0.61e+05 & 1.92e+05 & $\pm$ 0.68e+05 & 2.6e+05 & $\pm$ 0.62e+05 & -\\
$v2_3$ & 4.08e+05 & $\pm$ 0.62e+05 & 3.76e+05 & $\pm$ 0.67e+05 & 3.16e+05 & $\pm$ 0.61e+05 & -\\
$r1$ & 123.2 & $\pm$  3.5 & 122.8 & $\pm$  3.5 & 123.1 & $\pm$  3.5 & -\\
$r2$ & 5.874 & $\pm$0.011 & 5.876 & $\pm$0.011 & 5.876 & $\pm$0.011 & -\\
$r3$ & -0.087 & $\pm$0.028 & -0.112 & $\pm$0.028 & -0.112 & $\pm$0.028 & -\\
scale & -0.0024748 & $\pm$0.0000061 & -0.0024749 & $\pm$0.0000060 & -0.0024749 & $\pm$0.0000060 & -\\
$p_{u1}$ & 0 & - &  -1.1 & $\pm$  0.9 & 0 & - & -\\
$p_{u2}$ & 0 & - &  -2.8 & $\pm$  1.1 & 0 & - & -\\
$a$ & 0.01421 & $\pm$0.00040 & 0.01422 & $\pm$0.00040 & 0.01419 & $\pm$0.00040 & AU\\
$\xi$ & 0.436 & $\pm$0.050 & 0.417 & $\pm$0.053 &   2.2 & $\pm$  0.5 & -\\
$T_n$ &  1290 & $\pm$   39 &  1070 & $\pm$   62 &   22 & $^{+230}_{-22}$ & K\\
$\Delta_T$ &   699 & $\pm$   53 &   828 & $\pm$   92 &  2870 & $\pm$  430 & K\\
$T_s$ &  4530 & $\pm$  120 &  4530 & $\pm$  120 &  4520 & $\pm$  120 & K\\
albedo & 0 & - & 0 & - & 0.315 & $\pm$0.019 & -\\
\end{tabular}
\end{center}
\label{tab:results}
}
\end{table*}
\end{center}

\subsubsection{Spectrally-resolved phase curves}

Spectrally resolving the phase offset in principle allows information on the heat transport as a function of depth. To test for this effect we break the data down into 15 spectral bins and repeat the MCMC analysis described for the white light data, without allowing the additional radial freedom from limb darkening. We fix the system parameters to those found in the white light fit. We report the measurement of $\xi$ parameter as a function of wavelength in Table \ref{tab:specresults}.

\begin{table}{
			\caption{The best fitting value of $\xi$ as a function of wavelength. The value is poorly constrained compared to the white-light curve due to the lower signal to noise, and in several channels no significant value is detected.}
			\begin{center}
				\begin{tabular}{c c}
					\hline
					\hline
					Central wavelength &   log10(xi)  \\
					(microns) & \\
					\hline
1.1425 & -0.32 $\pm$ 0.27 \\
1.1775 & -0.51 $\pm$ 0.32 \\
1.2125 & -0.00 $\pm$ 0.18 \\
1.2475 & -0.20 $\pm$ 0.19 \\
1.2825 & -0.74 $\pm$ 0.39 \\
1.3175 & -0.32 $\pm$ 0.19 \\
1.3525 & -0.82 $\pm$ 0.62 \\
1.3875 & -0.07 $\pm$ 0.20 \\
1.4225 & -0.09 $\pm$ 0.34 \\
1.4575 & -0.71 $\pm$ 0.31 \\
1.4925 & -0.42 $\pm$ 0.17 \\
1.5275 & -0.76 $\pm$ 0.37 \\
1.5625 & -0.43 $\pm$ 0.14 \\
1.5975 & -0.66 $\pm$ 0.25 \\
1.6325 & -0.59 $\pm$ 0.19 \\
\end{tabular}
\end{center}
\label{tab:specresults}
}
\end{table}

\begin{figure}
	\begin{center}
		\includegraphics[width=\columnwidth]{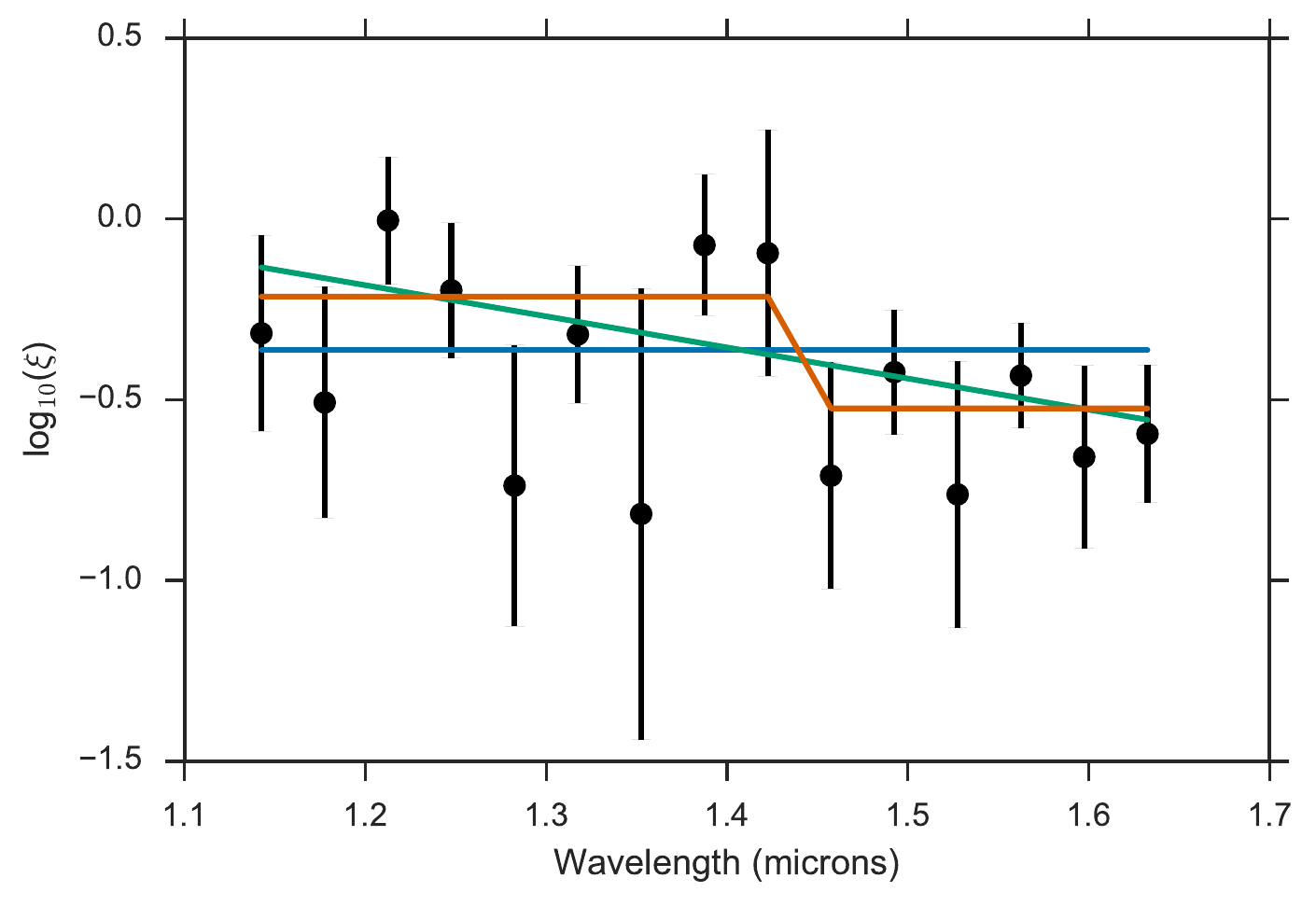}
		\caption{The values of log($\xi$) as a function of wavelength. Several simple models are fit to the data to test for wavelength dependence. A flat line in blue, a gradient model in green and a step function in orange.}
		\label{fig:best_fit_flux}
	\end{center}
\end{figure}

We test if there is a significant dependence of the value of $\xi$ on the wavelength in microns by attempting to fit several simple models to the data with a Levenberg Marquardt least squares technique. We choose to fit using the logarithm of the $\xi$ value, since the best fitting value is seen to vary by several orders of magnitude across spectral channels. 

The null model, which assumes there is no dependence on wavelength is simply a flat line. This model gave an acceptable fit, with a $\chi^2$ of 15.1 ($\chi^2_R$ of 1.16) with 14 DOF with a best fitting average value of -0.36 - which is consistent with the white light value. We fit a simple gradient model to test for linear trends in the data. The best fitting model had a $g$ -0.86 micron$^-1$ and a value of $c$ of 0.85 and gave a $\chi^2$ of 8.8 ($\chi^2_R$ 0.67) with 13 DOF. We also test a function approximating a step function, this form might be expected since the wavelength region covers the water band; values inside the water band will probe lower pressures than those outside, due to the increased optical depth. The function has three parameters: $\delta_{out}$, which is the average value outside of the step function, $\delta_{in}$ which is the value on the other side of the step, and $\gamma$, which is the wavelength which separates the two sides of the step function. The best fitting model was found to have $\gamma$ of 1.4225 microns, which is close to what would be expected for a water features, $\delta_{out}$ of -0.52 and $\delta_{in}$ of -0.21. This model gave a $\chi^2$ of 8.0 ($\chi^2_R$ of 0.62) with 12 DOF.

Using a Bayesian Information Criterion, the improvement of $\chi^2$ is marginal for the gradient model and the step function model, and not enough to constitute significant evidence in favour of either model, so we prefer the interpretation of a flat line. While we find that there is weak evidence for a trend with wavelength in the data, the flat line model provides a satisfactory fit to the data. A full treatment would require calculating the pressure contribution function for each wavelength as was done in \citet{Stevenson2014}, which is beyond the scope of this work.

\section{Conclusions}\label{sec:conclusions}

In this paper we have introduced \textsc{spiderman}, an open source code for modelling phase curves and secondary eclipses with arbitrary brightness distributions. The code is modular and able to use a variety of different brightness distributions.

The C-based algorithm analytically calculates the area occulted of radial segments on the planet's surface, and is designed to run rapidly to enable statistical techniques such as MCMCs to run quickly.

We demonstrate the use of \textsc{spiderman} by applying the code to the case of WASP-43b. We use a physically motivated analytical model presented by \citet{Zhang2016} which captures the major features of exoplanet phase curves. We find that the model provides a good fit to the data, but overpredicts the nightside flux compared to non-informative models. We hypothesise that this may be due to the presence of clouds on the planet's nightside, or alternatively, that the dayside possesses a non-zero near infrared albedo.

\begin{figure*}
\begin{center}
\includegraphics[width=\columnwidth]{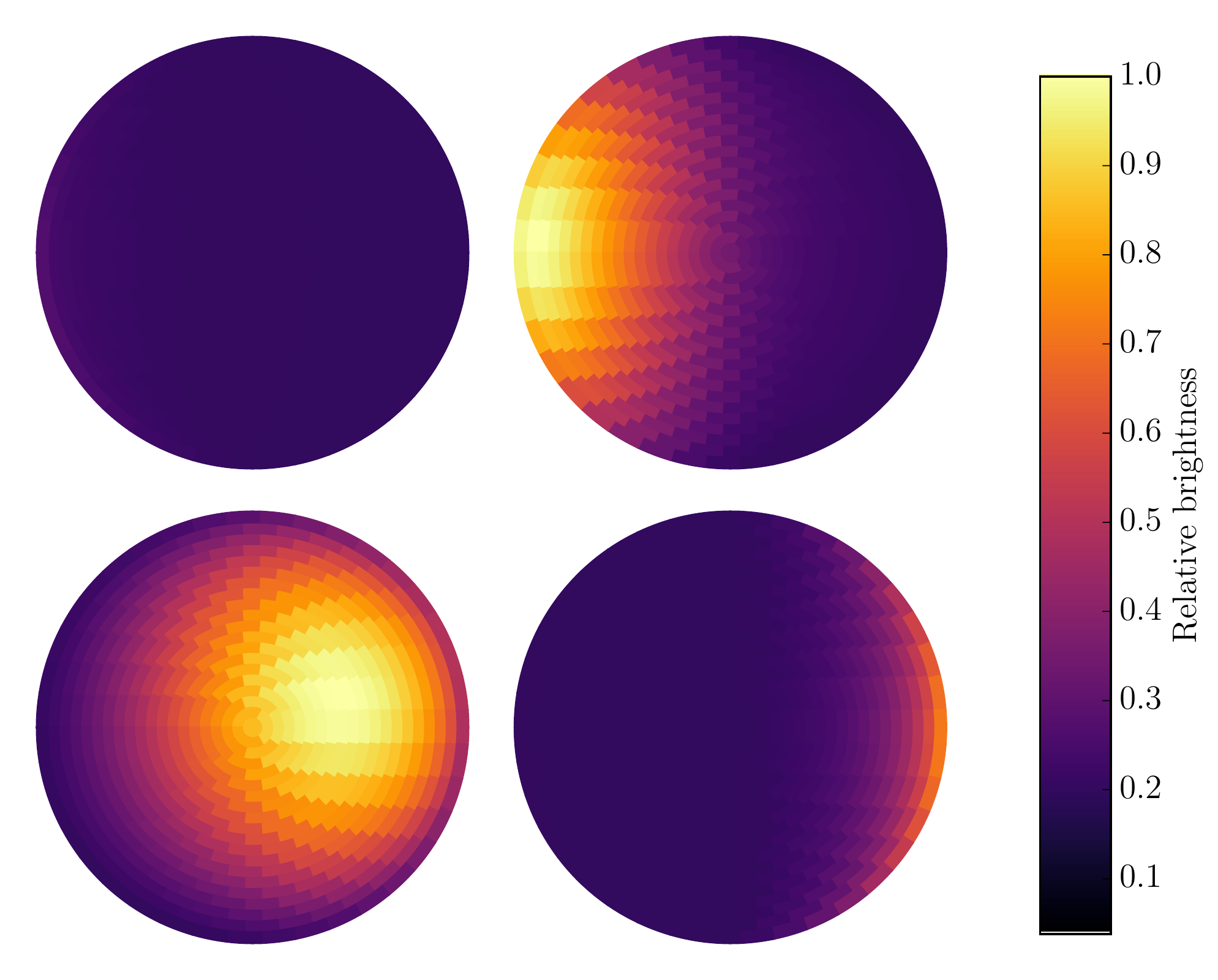}
\includegraphics[width=\columnwidth]{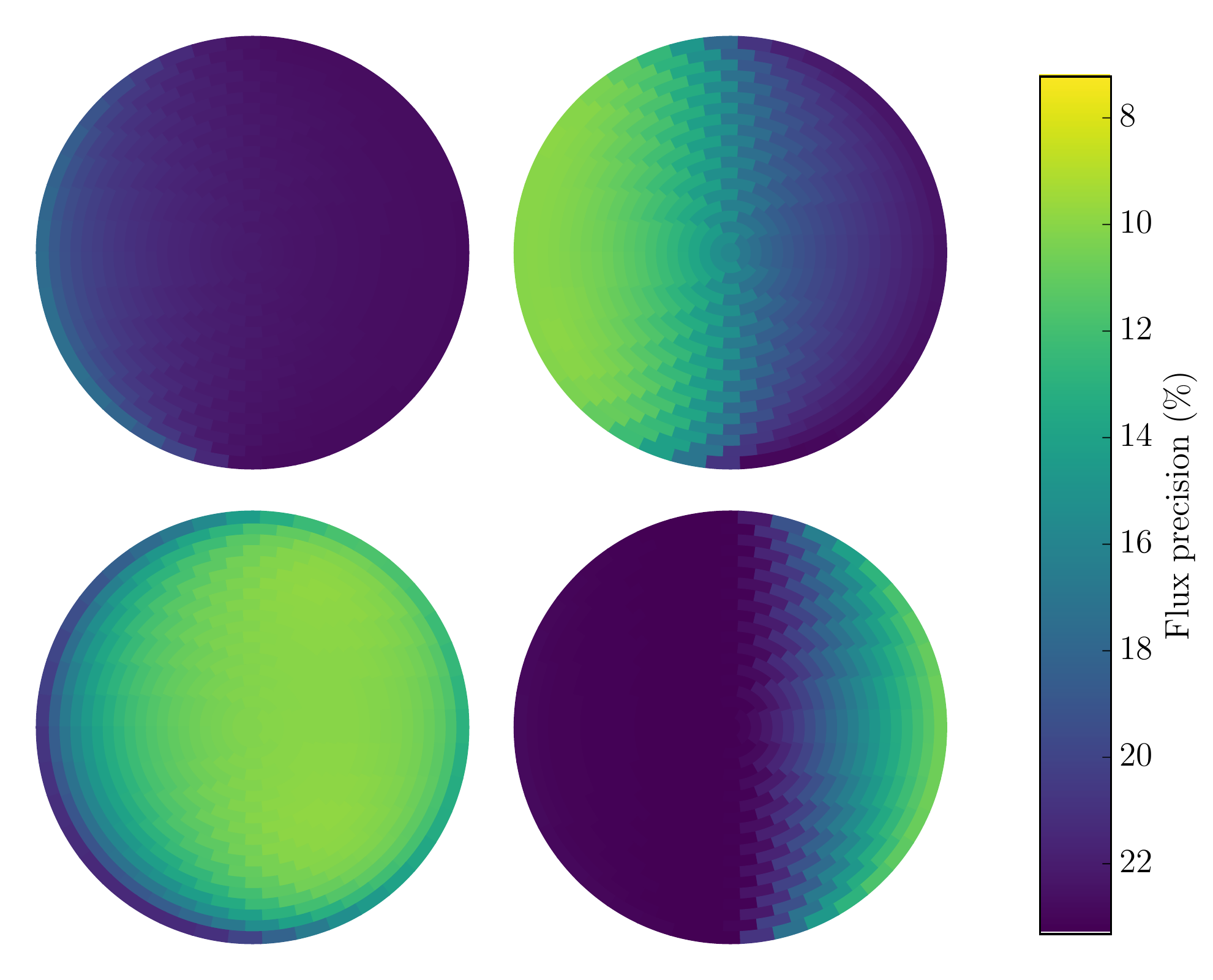}
\caption{Left: The best fitting flux distribution for the model where the planetary limb darkening was switched off at four phases. From top left clockwise the phases are 0, 0.25, 0.75 and 0.5. Right: The respective constraints on the brightness distribution.}
\label{fig:best_fit_flux}
\end{center}
\end{figure*}

\section*{Acknowledgements}

T.L. is supported by STFC consolidated grant (ST/P005586/1). Much of this work was carried out at the kavli summer program in physics 2016. T.L extends his gratitude to the program organisers, in particular Jonathan Fortney and Pascale Garaud. The authors thank Xi Zhang, Emily Rauscher, Nick Cowan, Jacob Arcangeli, and Jean-Michel D\'esert for useful discussions that led to improvements to \textsc{spiderman}.




\bibliographystyle{mnras}
\bibliography{Mendeley}



\newpage
\appendix

\section{All collision cases}

\subsection*{Case 2: zero collisions}
The trivial case where the segment is either entirely blocked, or entirely unobscured. If the distance from the center of the segment to the center of the star is less than the stellar radius then the entire area is added, otherwise nothing is.

\subsection*{Case 3: Inner and outer circles crossed once}
The case where both the inner and outer circle segments are both intersected by the disc of the star, and the straight edges are untouched. In this case the total area is calculated as follows
\begin{equation} \label{eq:inner_outer}
A_{tot} = A_1 + A_2 + A_3 - A_4
\end{equation}
$A_1$ is the circle segment of the stellar disc bounded by the intersection points with the inner and outer circles. $A_2$ is the circle segment of the outer circle edge bounded by the intersection point with the stellar disc and the outer corner that is blocked by the stellar disc. $A_3$ is the quadrilateral defined by the two intersection points and the two corners blocked by the stellar disc. $A_4$ is the circle segment of the inner circle edge bounded by the intersection point with the stellar disc and the inner corner that is blocked by the star. 

\subsection*{Case 4: Inner and outer circles crossed once, one straight edge crossed twice.}

\begin{equation} \label{eq:inner_outer}
A_{tot} = A_1 + A_2 + A_3 - A_4 - A_5
\end{equation}

This is calculated the same as Case 2, but an additional area element is subtracted, $A_5$, which is the circle segment formed by the disc of the star and the two intersection points with the straight edge.

\subsection*{Case 5: inner edge crossed twice, both straight edges crossed.}

There are two subcases; if the inner edges are covered by the star, then:

\begin{equation} \label{eq:inner_outer}
A_{tot} = A_R - (A_1 + A_2 + A_3 - A_4 - A_5)
\end{equation}

$A_R$ is the total area of the region.

otherwise, if the inner edges are outside the disc of the star:

\begin{equation} \label{eq:inner_outer}
A_{tot} = A_R - (A_1 + A_2 - A_3 - A_4 - A_5 - A_6)
\end{equation}

$A_R$ is the total area of the region. $A_1$ is the area of the triangle formed by the two corners on one straight edge, and the contact point with the inner edge that is closest to them, $A_2$ is the same for the other side. A$_3$ and A$_4$ are the circle segments of the inner edge, each formed by an inner corner and the intersection of the inner edge that is closest to them. A$_5$ and A$_6$ are the circle segments of the stellar disc, each formed by an intersection of a straight edge and the intersection of the inner edge that is closest to them.

\subsection*{Case 6: A straight edge crossed twice}

In this case, only a single area needs to be calculated, the area of a circle segment defined by the stellar disc, and the two intersection points

\subsection*{Case 7: Inner edge and one straight edge}

Similar to Case 2, this case has two subcases, one where there are three blocked corners, and one where one corner is blocked. For the case with one corner inside:

\begin{equation} \label{eq:inner_outer}
A_{tot} = A_1 + A_2 - A_3
\end{equation}

$A_1$ is the triangle defined by the corner that is blocked by the disc and the two intersection points. $A_2$ is the circle segment of the star disc defined by the two intersection points. $A_3$ is the segment area of the inner edge defined by the intersection point and the corner that is blocked by the stellar disc.

For the case with three corners inside:

\begin{equation} \label{eq:inner_outer}
A_{tot} = A_R - (A_1 - A_2 - A_3)
\end{equation}

$A_R$ is the total area of the region. $A_1$ is the segment of the inner edge defined by the intersection point and the corner not blocked by the star. $A_2$ is the segment of the star defined by the intersection and the corner not blocked by the star. $A_3$ is the triangle defined by the corner not blocked by the star and the two intersection points.

\subsection*{Case 8: Outer edge crossed twice}

If no corners are covered by the star, then this is simply the sum of two circle segments, for the outer edge and the star, defined by the two interestion points.

If all corners are covered by the star, then:

\begin{equation} \label{eq:inner_outer}
A_{tot} = A_R - (A_1 - A_2)
\end{equation}

Where $A_1$ is the segment of the outer edge defined by the two interestion points, and $A_2$ is the segment of the stellar disc defined by the two interestion points. 

\subsection*{Case 9: Both straight edges crossed}

This is divided further into two subcases, if the two outer corners are covered by the star, then:

\begin{equation} \label{eq:inner_outer}
A_{tot} = A_1 + A_2 + A_3
\end{equation}

$A_1$ is the quadrilateral defined by the intersection points and the outer corners. $A_2$ is the circle segment defined by the stellar disc and the intersection points. $A_3$ is the circle segment defined by the outer edge and the outer corners.

Otherwise, if the two inner corners are covered:

\begin{equation} \label{eq:inner_outer}
A_{tot} = A_1 + A_2 - A_3
\end{equation}

$A_1$ is the quadrilateral defined by the intersection points and the inner edges. $A_2$ is the circle segment defined by the stellar disc and the intersection points. $A_3$ is the circle segment defined by the inner edge and the inner corners.

\subsection*{Case 10: 2 inner, one outer, one edge}

This case has two subcases. In the first case the center of the star is external to the outer circle edge of the segment, the area is:

\begin{equation} \label{eq:inner_outer}
A_{tot} = A_1 + A_2 + A_3 - A_4
\end{equation}

Where $A_1$ is the triangle formed by the intersection with the outer circle edge and the two segment corners blocked by the star, $A_2$ is the circle segment of the formed by the interestion with the outer circle edge and the outer corner which is blocked by the star, $A_3$ is the circle segment of the star formed by the interestion with the outer circle and the intersection with the edge and $A_4$ is the area of the intersection of the circles describing the inner circle edge and the disc of the star.

in the second case the center of the star is interior to the inner circle edge, and:
\begin{equation} \label{eq:inner_outer}
A_{tot} = (A_1 + A_2 +A_3 - A_4) + (A_5+A_6-A_7)
\end{equation}
$A_1$ is the area of the quadrilateral definted by the corners of the straight edge that is fully covered, (hereafter edge 1), the interesection with the outer circle, and the closest intersection with the inner edge (hereafter $c_1$). $A_2$ is the segment of the outer circle formed by the intersection with the outer circle and the outer corner of edge 1. $A_3$ is the circle segment of the star formed by the interesction with the outer circle and $c_1$. $A_4$ is the circle segment of the inner circle formed by the inner corner of edge 1 and $c_1$. $A_5$ is the triangle formed by the intersection with \emph{edge 2}, the inner corner of edge 2 and the remaining inner circle intersection, $c_2$. $A_6$ is the circle segment of the star defined by the intersection with edge 2 and $c_2$. $A_7$ is the circle segment of the inner circle defined by the inner corner of edge 2 and $c_2$.

\subsection*{Case 11: 2 inner, two outer}

This case only occurs when the star is smaller than the planet.

\begin{equation} \label{eq:inner_outer}
A_{tot} = A_1 + A_2 + A_3 + A_4 - A_5
\end{equation}

$A_1$ is the quadrilateral formed by the 4 intersection points. $A_2$ and $A_3$ are the circle segments of the stellar disc defined by the two pairs of inner and outer intersection points. $A_4$ is the circle segment of the outer edge formed by the two intersection points on the outer edge. $A_5$ is the circle segment of the outer edge formed by the two intersection points on the inner edge.

\subsection*{Case 12: tangent solutions}

The remaining posible cases are all tangent solutions, where the disc of the star touches the boundary of the region exactly once, and no area is blocked.	



\bsp	
\label{lastpage}
\end{document}